\documentclass[10pt, a4paper]{article}

\usepackage{graphicx}
\usepackage[authoryear,sort,round]{natbib}
\usepackage{amsmath,authblk,adjustbox,caption,subfig}
\usepackage[hidelinks]{hyperref}

\def\sym#1{\ifmmode^{#1}\else\(^{#1}\)\fi}
\newcommand{\myFiguresSize}{.74\columnwidth}

\graphicspath{{figures/}}

\usepackage{color}

\begin{document}

\title{The Indirect Effects of FDI on Trade: \\ A Network Perspective\thanks{The authors acknowledge the funding from the Italian Ministry of Education, University and Research (MIUR) through the National Research Program of Italy (PNR), the CRISIS Lab project, and the project ``The global virtual-water network: social, economic, and environmental implications'' (FIRB - RBFR12BA3Y).
They would like to acknowledge helpful comments from Rene Belderbos, Giorgio Fagiolo, and Enrico De Angelis and from the participants in the ECCS 2014 conference in Lucca (Italy). Our special thanks are due to Armando Rungi for his insightful suggestions and his generous support of our data preparation. 
M.R. acknowledges the funding from the Multiplex FP7 project (Foundational Research on MULTIlevel comPLEX networks and systems). 
}}

\author[1]{\small Paolo Sgrignoli}
\author[2]{\small Rodolfo Metulini}
\author[1]{\small Zhen Zhu}
\author[1,3]{\small Massimo Riccaboni}
\affil[1]{\small  LIME, IMT School for Advanced Studies Lucca, Lucca, Italy}
\affil[2]{\small  Department of Economics and Management, University of Brescia, Brescia, Italy} 
\affil[3]{\small  DMSI, KU Leuven, Leuven, Belgium}

\date{}
\maketitle

\begin{abstract}
\noindent  The relationship between international trade and foreign direct investment (FDI) is one of the main features of globalization. 
In this paper we investigate the effects of FDI on trade from a network perspective, 
since FDI takes not only direct but also indirect channels from origin to destination countries because of firms' incentive to reduce tax burden, to minimize coordination costs, and to break barriers to market entry. 
We use a unique data set of international corporate control as a measure of stock FDI 
to construct a corporate control network (CCN) where the nodes are the countries and the edges are the corporate control relationships. 
Based on the CCN, the network measures, i.e., the shortest path length and the communicability, are computed to capture the indirect channel of FDI.
Empirically we find that corporate control has a positive effect on trade both directly and indirectly. The result is robust with different specifications and estimation strategies. Hence, our paper provides strong empirical evidence of the indirect effects of FDI on trade. Moreover, we identify a number of interplaying factors such as regional trade agreements and the region of Asia. 
We also find that the indirect effects are more pronounced for 
manufacturing sectors than for primary sectors such as oil extraction
and agriculture. 
\end{abstract}



\newpage

\section{Introduction}\label{sec:intro}

The relationship between international trade and foreign direct investment (FDI), which is one of the main features of globalization, is complex and it is not limited to the issue of whether they are complementary\footnote{The reasoning behind is that the bilateral trade will be decreased if the bilateral FDI is horizontal or will be increased if the bilateral FDI is vertical \citep{markusen1997trade,markusen2004multinational,markusen2003general}.} or not \citep{fontagne1999foreign}.

Previous studies on the effects of FDI on trade are by and large confined to a two-country setting, where bilateral trade is solely determined by the characteristics of the two countries considered. 
However, the empirical tests based on the two-country setting have concluded with mixed results. For example, relying on a cross-sectional firm survey data set from the 1970s in the United States, \citet{lipsey1984foreign} find that a US firm's outward FDI to a foreign area is positively associated with its exports to that foreign area. 
Based on a panel of China's bilateral data with 19 foreign areas during 1984-1998, 
\citet{liu2001causal} also show that the inward FDI from a foreign area to China induces China's exports to that foreign area. 
Conversely, \citet{belderbos1998tariff} find a negative relationship between Japanese electronics firms' exports to Europe and their investment in Europe in the late 1980s when Europe adopted a strict antidumping policy. They also find that more trade is created if the investment is related to global value chains (GVCs). 
Moreover, \citet{blonigen2001search} finds a mixed relationship between affiliate production and exports of Japanese automobile products in the United States from the late 1970s to the early 1990s. 
Finally, \citet{amiti2003investment} run a gravity model for every year in a panel of bilateral data between 36 countries during 1986-1994 and find that FDI has a positive (or negative) effect on trade when countries are different (or similar) in terms of factor endowments and when trade costs are low (or high). 

\begin{figure}[!t]
	\centering
	\caption{Direct and indirect costs.} 
	\includegraphics[width=\myFiguresSize]{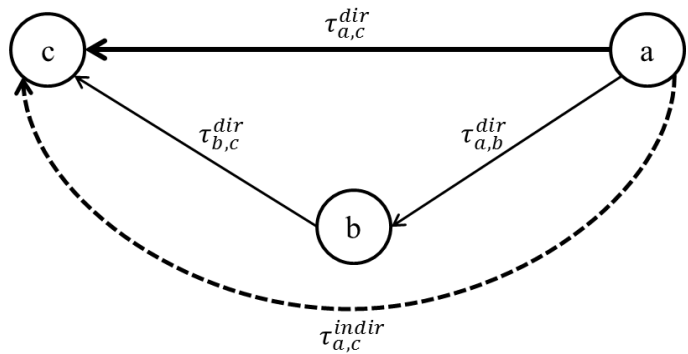}
	\label{fig:cost_example}
\end{figure}

Our paper investigates the effects of FDI on trade beyond the two-country setting. Before examining its effects on bilateral trade, we quantify FDI from a network perspective and consider both the direct and indirect channels of FDI.   
The indirect channels of FDI can be explained by firms' cost-minimization
strategies.  
Figure \ref{fig:cost_example} shows an imaginary three-country example where country $a$ intends to conduct outward FDI to country $c$.
The costs of doing so directly is $\tau_{a,c}^{dir}$ and indirectly via country $b$ is $\tau_{a,c}^{indir}$, which is a combination of 
$\tau_{a,b}^{dir}$ and $\tau_{b,c}^{dir}$.\footnote{For simplicity, we
	only consider a single intermediate country $b$. But in principle, the 
number of intermediate countries can be greater than 1.}  
Note that it is possible that $\tau_{a,c}^{indir}<\tau_{a,c}^{dir}$ and 
country $a$ prefers indirect FDI if this is the case.    
Below we highlight some concrete examples of costs that firms may take into
consideration when undertaking indirect FDI.

First, firms may prefer indirect FDI so as to reduce their tax burden. 
Due to the varying availability of tax and investment treaties across countries, a parent company may locate an affiliate in an intermediate country to control another affiliate in the final destination country so as to receive the most favorable tax and investment treatment in the host countries by profit shifting \citep{hines1994fiscal,benassy2005does,van2014ranking,gumpert2016multinational}.
For example, as a well-connected country in terms of tax and investment treaties, 
the Netherlands is the world's largest pass-through country for approximately 1600 billion euros of FDI in 2009 \citep{weyzig2013tax}. 
\citet{mintz2010indirect} also find that in 2012 about 15 percent of outbound German FDI affiliates are
held via an intermediate firm in a third country, with the Netherlands
as the most important location of the so-called conduit entities. 
Other famous examples include the FDI ``round-tripping'' through Hong Kong to China and through Mauritius to India \citep{wei2005china}.

Second, another driver of indirect FDI is to reduce coordination costs. 
That is, a parent company may locate an affiliate in a intermediate foreign country if it is geographically, politically, or culturally closer to the final destination country so as to achieve more effective coordination \citep{kalotay2012indirect}. 
For example, Russia indirectly invests in Central and Eastern European countries through Cyprus, taking advantage of the latter's accession in the European Union \citep{pelto2004cyp}. 
Note that this driver may, to some extent, overlap with the above driver of tax and investment treaties. 
The fact that Hong Kong and Mauritius serve as the gateways of inward FDI to China and India respectively is also due to the geographical and cultural closeness. 

Third, indirect FDI can be a dynamic process and can be considered as
a growth strategy to break barriers to market
entry. 
In practice, facing high startup costs if directly holding an affiliate 
in the final destination country, multinational companies often divide a target region into several
clusters of countries and pursue a regional management structure, 
where regional management centers are established first with strategic and 
operational roles in each cluster \citep{amann2014clusters}. 
Moreover, FDI can be a consequence of experience accumulated and can be 
conceived as a sequential process when crossing multiple country borders \citep{kogut1983foreign}.

Our paper highlights the importance of indirect effects and it is therefore related to the recent literature of FDI and trade investigating the third-country effects \citep{baltagi2007estimating,blonigen2007fdi,garretsen2009fdi}. 
Besides the simple dichotomy of horizontal and vertical FDI, the mixed nature of FDI
has been noted in the literature
and new terms such as ``complex
FDI'' and ``networked FDI'' have been created to account for more structured forms of FDI such as export platforms and production networks
\citep{yeaple2003complex,ekholm2007export,baldwin2014networked}.
However, our paper differs from these studies in at least two aspects. First, while they study the determinants of FDI, we are interested in the consequences of FDI on trade.\footnote{Even more recently, \citep{park2015modes} study the effects of FDI on trade by considering the third-country effects. They use spatial econometrics and focus on the inward FDI to China, whereas we use network analysis and a cross-country data set.} 
Second, to capture the third-country effects, they use spatial econometrics whereas we use the tools of network analysis.\footnote{In fact, we perform
	a spatial econometrics exercise with our data set and find a significant third-country effect, which motivates us to study the indirect effects
	of FDI on trade from a network perspective. The spatial econometrics
result is available upon request.}

To take into account both the direct and indirect channels of FDI, we use a unique data set of international corporate control (as a measure of stock FDI) covering almost all countries over the world,  
while previous results are often based on a small sample of countries or case studies due to the paucity of FDI data.  

Furthermore, we  construct a corporate control network where the nodes are the countries and the edges (both directed and weighted) are the corporate control relationships. 
Based on the corporate control network, the shortest path length (which can be either direct or indirect) and the communicability (which is an overall measure of ``communication'' between nodes) between each pair of countries are computed.
The shortest path length (or the communicability) complements (or substitutes) the direct corporate control intensity and together they provide a more complete accounting of the effects of FDI on trade.

Then we find, using the Heckman two-stage (H2S) gravity model, a positive effect of FDI on trade both for the direct corporate control intensity and the indirect measures, i.e., the shortest path length and the communicability. Therefore our paper provides strong empirical evidence of the importance of indirect effects of FDI on trade. 
We also identify a number of interplaying factors, including regional trade agreements (RTAs) and the region of Asia. 
We also find that, compared with primary sectors such as oil extraction
and agriculture, manufacturing sectors have more pronounced indirect
effects of FDI on trade. 

The remainder of this paper is structured as follows: Section \ref{sec:networks-comparison} introduces FDI and trade networks and describes the network measures of indirect effects; Section \ref{sec:data-and-definitions} describes our data set and provides some exploratory analysis; Section \ref{sec:economic-and-econometric-approach} specifies our econometric methodology while Section \ref{sec:results} presents our main results; finally, Section \ref{sec:conclusions} concludes the paper.

\section{Networks of FDI and Trade}\label{sec:networks-comparison}

\subsection{Global Systems as Networks}

There is a significant body of literature developed recently in studying economic phenomena from a network perspective. For example, economic systems such as international trade \citep{frankel2002estimate,glick2002does,serrano2003topology,garlaschelli2005structure,fagiolo2009world,de2011world,reyes2014regional} and corporate control \citep{head2008fdi,vitali2011network,Altomonte2013}
can be considered as networks and their network properties can be used to understand other economic variables \citep{Schiavo2010,riccaboni2013global,Fagiolo2014,Sgrignoli2015,ferrier2016technology}. 

For the world system of either trade or corporate control, we can construct a network, where we identify countries as nodes and interaction channels between them as edges. 
For simplicity, we call them the world trade web (WTW) and the corporate control network (CCN) respectively. As a result, we have both networks composed of $n$ nodes (countries). 
The two networks are based on two kinds of weighted directed edges, i.e., two $n\times n$ adjacency matrices, one corresponding to trade flows ($T$) and the other to corporate control links ($C$). 
The generic element of $T$ (or $C$) represents the value of exports $T_{ij}$ (or the number of corporate control ties $C_{ij}$) from country $i$ to country $j$.

\subsection{Network Measures of Indirect Effects}

As discussed above, factors such as GVCs, tax and investment treaties, and corporate strategies, allow the indirect effects of FDI on trade between countries.  
To capture the indirect effects of FDI on trade, we use two types of network measures based on the CCN. 
One is the shortest path length \citep{brandes2001faster,newman2001scientific,opsahl2010node}. Recall that $C_{ij}$ is the edge weight from country $i$ to country $j$ on the CCN. Define the direct path length from $i$ to $j$ as $P_{ij}^{dir}=\frac{1}{C_{ij}^\alpha}$, where $\alpha\geq 0$.
The indirect path length from $i$ to $j$ is then computed by adding up the direct path lengths from $i$ to $j$: 
\begin{equation} \label{eqn:indirPath}
	P_{ij}^{indir}=P_{ih}^{dir}+\dots+P_{gj}^{dir}=\frac{1}{C_{ih}^\alpha}+\dots+\frac{1}{C_{gj}^\alpha}
\end{equation}
where $\alpha\geq 0$ and $h,\dots,g$ are the intermediate countries between $i$ and $j$. We follow \citep{ferrier2016technology} and choose $\alpha=1$ as our benchmark\footnote{We also discount the importance of indirect links with respect to direct ones by choosing $\alpha=0.5$ and the main regression results stay the same. See Tables \ref{tab:res_ups_dist_newAlpha} and \ref{tab:res_rta_newAlpha} in the appendix.}.    

\begin{figure}[!t]
	\centering
	\caption{Direct and indirect paths.} 
	\includegraphics[width=\myFiguresSize]{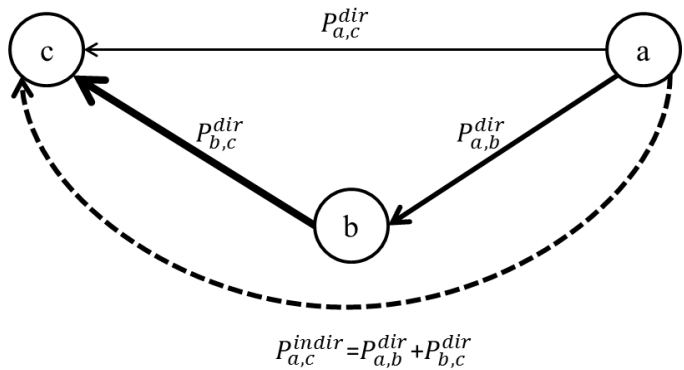}
	\label{fig:spl_example}
\end{figure}

In the same spirit of Figure \ref{fig:cost_example}, Figure \ref{fig:spl_example} shows the difference between the direct and indirect path lengths in a three-country example. Note that the indirect path $P_{a,c}^{indir}$ may be shorter than the direct path $P_{a,c}^{dir}$ according to Equation \ref{eqn:indirPath}.\footnote{Again, for simplicity, we only consider a single
	intermediate country $b$. But as shown in Equation \ref{eqn:indirPath}, 
the number of intermediate countries can be greater than 1.}

Finally, let $\boldsymbol{\mathrm{P}}_{ij}^{indir}$ be the set of all possible indirect paths from $i$ to $j$, the shortest path length from $i$ to $j$ is defined as:
\begin{equation}\label{eqn:spl}
	{spl}_{ij}=\min\left\{P_{ij}^{dir},\boldsymbol{\mathrm{P}}_{ij}^{indir}\right\}
\end{equation}

Note that $P_{ij}^{dir}$ may not exist from $i$ to $j$ and that there may be no paths from $i$ to $j$ at all (in this case ${spl}_{ij}=\infty$ and we treat it as a missing value).

The other network measure to capture the indirect effects is the communicability, which takes into account not only the shortest path but also all the other walks that connect one node to another \citep{estrada2008communicability}. 

Let $A$ be the adjacency matrix of an undirected and unweighted network
and $A_{ij}=A_{ji}$ equals 1 if there is an edge between $i$ and $j$ and otherwise equals 0. 
A well-known property of $A$ is that the $(i,j)$ entry of the $s$th power of $A$, $(A^s)_{ij}$, returns the number of walks of length $s$
starting at $i$ and ending at $j$. A walk of length $s$ is
defined as a sequence of nodes $v_0,v_1,\dots,v_{s-1},v_s$ such that, 
for each $i=1,2,\dots,s$, a link exists from $v_{i-1}$ to $v_i$. Note that
the nodes involved in a walk are not necessarily different from each 
other (i.e., some nodes may be revisited).

Then the communicability is defined as:
\begin{equation} \label{equ:cmb0}
	{cmb}_{ij}=\sum_{s=0}^\infty \frac{(A^s)_{ij}}{s!}
\end{equation}
where $s$ is used to discount the number of walks of length $s$ because
less importance should be given to longer walks.  

The eigenvalues of $A$ in the non-increasing order 
$\lambda_1\geq\lambda_2\geq\dots\geq\lambda_n$ are also called the 
spectrum of the graph. And Equation \ref{equ:cmb0} can be rewritten in 
terms of the graph spectrum:
\begin{equation} \label{equ:cmb}
	{cmb}_{ij}=\sum_{k=1}^n \phi_k(i)\phi_k(j)e^{\lambda_k}
\end{equation}
where $\phi_k(i)$ is the $i$th element of the $k$th orthonormal eigenvector of the adjacency matrix associated with the eigenvalue $\lambda_k$. 

Note that, unlike the shortest path length, the communicability is based on an undirected and unweighted version of the original network, i.e., ${cmb}_{ij}={cmb}_{ji}$. 
While the shortest path may be either direct or indirect, the communicability takes into account all possible walks (including paths). 
As a result, it provides a simplified but comprehensive measure of the level of ``communication'' between nodes.   
In our econometric analysis below, we use the shortest path length as the main
network measure and use the communicability as a robustness check.

\section{Data and Descriptive Analysis}\label{sec:data-and-definitions}

\subsection{Data}

In our empirical analysis we use corporate control data as a measure of stock FDI. Our corporate control data is taken from the ORBIS database compiled by Bureau Van Dijk for the year 2010\footnote{A related work using the ORBIS database is \citep{Altomonte2013}. Unlike theirs, our data usage is restricted to the cross-country control links.}. 
We only consider the cross-country ownership relationships and a control is assumed if the parent company holds the voting rights majority (50.01\%) of the affiliate in another country. 
The data thus indicates for each pair of countries the number of control links present between them for both directions.
The data we end up with contains $36,461$ ultimate parent multinational firms, controlling a total of $354,569$ affiliates in 209 countries, for the year 2010. In our data, parent firms located in OECD economies hold around 84\% of all the control links, the 63\% of which are still in an OECD country. The headquarters located in European Union countries, in particular, control 46\% of all affiliates, of which roughly three quarters are located outside the Union. 

Our trade data is taken from the BACI dataset \citep{CEPII:2010-23}, which originates from the data reported by over 150 countries to the United Nations Statistics Division (COMTRADE database) but also integrates new approaches to reconcile those reports, in order to have a single consistent figure for each bilateral flow. The version (Harmonized System 1996 or simply 
HS96) we use covers more than $200$ countries and $5000$ products, between 1998 and 2012.

Using the two data sets together, we obtain a data set of 191 countries (nodes) along with trade and corporate control relationships among them. Since the only year present in both datasets is 2010, we perform a cross-sectional analysis on that year.

We employ additional country-specific data such as real gross domestic product (GDP) per capita ($gdp$) and population ($pop$) from the World Bank. We also use bilateral country geographic, political, and socioeconomic data from the CEPII GeoDist dataset \citep{mayer2011notes}. The latter includes information about between-country geographical distance ($dist$\footnote{We employ the great-circle definition of country distances. Results do not change if we use alternative distance definitions.}), 
contiguity ($contig$, i.e., whether two countries share a border), 
colony relationship ($colony$, i.e., whether one of the two countries has ever been a colony of the other), 
whether two countries have ever been unified ($smctry$), and ethnical language commonality ($comlang$, i.e., if spoken by at least 9\% of population). 
We use the above variables to implement the gravity model in Section \ref{sec:results}.

\subsection{Exploratory Analysis}

\begin{figure}[!t]
	\caption{WTW and CCN in 2010. The figure plots the directed top 2.5\% edges by weight. Blue edges represent trade-only relations and red ones represent corporate-control-only, while green ones indicate the presence of both. The edge thickness is proportional to the log of edge weight. Node size is proportional to the log of population ($pop$), while color (from blue to red) is proportional to the log of real GDP per capita ($gdp$).}
	\label{fig:maps}
	\centering
	\includegraphics[width=\myFiguresSize]{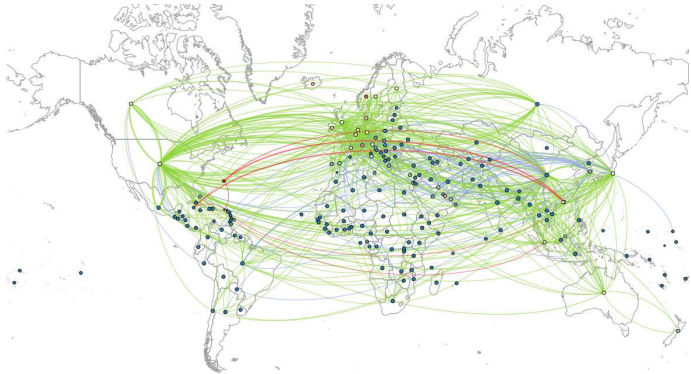}
\end{figure}

To give an idea of how the trade and FDI networks look like, in Figure \ref{fig:maps} we plot the top 2.5\%\footnote{We have also checked the thresholds of 1\% and 5\% and they convey the same information.} directed edges in terms of edge weight, according to the following criteria:
the edge color identifies whether the relationship is solely in trade (blue), solely in corporate control (red), or in both (green); 
the edge thickness is proportional to the log of edge weight\footnote{In the cases where both trade and corporate control are present (green edges), 
the weight is calculated as the mean of the two after normalization.}; and the node size and color are proportional, respectively, to the log of population and the log of real GDP per capita.

It is straigtforward to see that most of the significant connections are characterized by both trade and corporate control (green) and the most intensive interaction happens between Europe and the United States. Another interesting finding is that the corporate-control-only connections (red) are primarily associated with ``tax havens'' such as Bermuda and Cayman Islands.\footnote{Other ``tax havens'' identified by the red links include British Virgin Islands, Hong Kong, and Singapore. Note that there is a link between Bermuda and China and every other red link is between ``tax havens.''}    
Therefore, what Figure \ref{fig:maps} suggests is that FDI and trade are strongly correlated and that FDI has a preference over ``tax havens''.

\begin{figure}[!t]
	\centering
	\caption{WTW vs CCN edge weights for the aggregate level. The dot color (blue to red) is proportional to the log of $\frac{{gdp}_i \times {gdp}_j}{dist_{ij}}$. The dot size is proportional to the log of $\frac{{pop}_i \times {pop}_j}{dist_{ij}}$.}
	\includegraphics[width=\myFiguresSize]{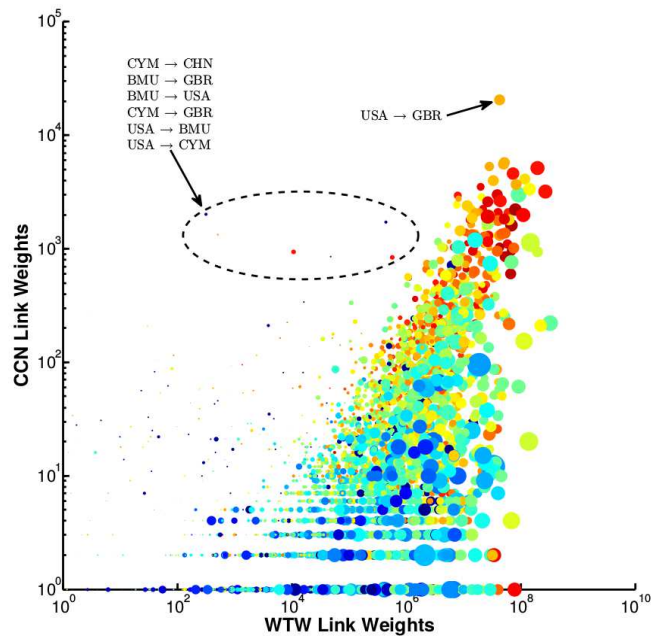}
	\label{fig:weights}
\end{figure}

As another exploratory analysis, Figure \ref{fig:weights} shows the log-log scatter plots of the WTW and the CCN edge weights. 
Each dot is an element in the space $(T_{ij}, C_{ij})$, i.e., the space of the two networks edge weights. The dot color is proportional to the log of $\frac{{gdp}_i \times {gdp}_j}{dist_{ij}}$ and the dot size is proportional to the log of $\frac{{pop}_i \times {pop}_j}{dist_{ij}}$.
The rationale behind this analysis resides in the well-known empirical success of the gravity model for FDI, but especially for trade, i.e., both goods and investments flows are well explained by a gravity-like equation involving country sizes (e.g., $gdp$ and $pop$) and geographical distance. 
In its simplest form, the gravity model of trade prescribes direct proportionality to countries' sizes and inverse proportionality to their geographical distance, i.e.,

\begin{equation}\label{eq:newton}
	F_{ij} \propto \frac{s_i s_j}{d_{ij}}\ ,
\end{equation} 
where $F_{ij}$ represents the flow between country $i$ and country $j$, $s_i$ and $s_j$ represent their respective sizes, and $d_{ij}$ is the geographical distance between the two.

If the gravity rules, one should expect that most of the variation in the cloud of points can be explained by larger country sizes and smaller distances. In our case $gdp$ (dot color) plays a more evident role as richer pairs of countries tend to be located in the north-east part of the plot, whereas  $pop$ (dot size) has less explanatory power.
Furthermore, Figure \ref{fig:weights} suggests a positive relationship between the edge weights in the two networks, as a high level of exports is in general associated with a high number of corporate control links.

It is also interesting to notice which the outlier edges are. We find again that most of them are ``tax havens,'' where there are intense incoming and outgoing corporate control links and relatively low flows of goods (in Figure \ref{fig:weights} we highlight only a few).

\section{Econometric Specifications}\label{sec:economic-and-econometric-approach}

As stated in Section \ref{sec:intro}, besides the direct effects of FDI on trade between countries, the indirect effects of FDI on trade are possible due to factors such as GVCs, tax and investment treaties, and corporate strategies.  
Therefore, when explaining trade, we introduce the network measures of the indirect effects. In particular, we either complement the direct corporate control intensity with the shortest path length or substitute it with the communicability. We do not include both the direct corporate control intensity and the communicability at the same time because the two variables are highly correlated\footnote{The reason why they are so correlated is that the communicability takes into account both the direct link and the indirect links and assigns the largest weight to the direct link. See Table \ref{tab:corr} for the correlation coefficients among the variables we used in the regressions.} in our data and would therefore produce biased estimations.
Therefore, we only include the communicability in the regressions to account for an overall measure of FDI relationships between countries. 

Moreover, what we expect to see is a strong correlation between corporate control links and trade flows in the two possible ways, i.e., both when the two relationships are in the same direction (corporate control export and trade export, e.g., a parent company exports inputs to its foreign affiliates) and when they happen in opposite directions (corporate control export and trade import, e.g., a parent company imports processed inputs from its foreign affiliates). 
We also intend to test the effects of a group of factors that may affect whether corporate control and trade are substitutes or complements, including regional trade agreements (RTAs) and the region of Asia.

An important feature of today's globalized and integrated economy is the proliferation of regional trade agreements (RTAs). According to the WTO data, the number of RTAs has risen from less than 100 in the early 1990s to over 300 today and more than half of the world trade is governed by at least one RTA \citep{WTO2011,Damuri2012}. While one might be tempted to think that trade is naturally of a global span, it is in fact very regional\footnote{The word ``regional'' here refers to big international economic blocks such as EU, NAFTA, and ASEAN.}. Without a widespread harmonization of trade and investment agreements \citep{WTO2013}, it is reasonable to suspect different interactions between trade and corporate control depending on whether an RTA is in place.

Finally, another important aspect to test is the special case of Asia. As one can observe in Figure \ref{fig:maps}, the trade relationships are particularly concentrated in Asia. Therefore, we want to test if being in Asia has any significant effects on trade and on the relation between corporate control and trade.

Now we turn to the econometric specifications. In the following analyses we use the Heckman two-step (H2S) \citep{Heckman1979,helpman2008estimating} gravity equation to model the relation between trade and corporate control (as a measure of stock FDI).

Since the seminal works by \citet{linnemann1966econometric} and \citet{tinbergen1962shaping}, the gravity model has been widely used in empirical studies on trade because of its excellent fit with the data \citep{frankel2002estimate, egger2002econometric}.
Moreover, a lot of effort has been put into refining it and giving it a consistent economic foundation \citep{bergstrand1985gravity, anderson2003gravity}. Generalizing Equation \ref{eq:newton} above, and applying a log transformation of both sides, we can write the formula in a linear multivariate form:
\begin{equation}
	\ln F_{ij} = \beta_0 + \beta_1 \ln s_i + \beta_2 \ln s_j + \beta_3 \ln d_{ij} + \ldots + \epsilon_{ij}
\end{equation} 
where $\epsilon_{ij}$ is the stochastic residual term, usually assumed to be i.i.d. and $\sim N(0, \sigma^2)$, and ``$\ldots$'' indicates the possibility to add more regressors (e.g., country- or edge-specific characteristics) to the model specification.

Empirically, the size variables ($s_i$ and $s_j$) typically include GDP per capita\footnote{Unlike in the previous exploratory analysis, where real GDP per capita is used, we use nominal GDP per capita in the regressions as the dependent variable is in current US dollars.}
($gdp$) and population ($pop$) while the impedance factors ($d_{ij}$) typically include geographical distance ($dist$). Since we focus on the effects of FDI on trade, besides the traditional variables of a gravity model, we control for both the direct corporate control intensity ($CC$) and the shortest path length ($spl$).

The two-step sample selection estimators are used to model bilateral trade in the presence of zero flows, as they allow us to remove the effects of the extensive margin of trade in order to correctly estimate the intensive margin effects, in contrast with other biased approaches which calculate coefficients that combine both the extensive and intensive margins \citep{hema13}. \citet{helpman2008estimating} also show that traditional estimates are biased and that most of the bias is not due to selection but rather to the omission of the extensive margin. 
Moreover, \citet{linders2006estimation} conclude that censored or truncated regression and replacement of zero flows with arbitrary numbers are not preferable as these approaches may yield misleading results and they rely on ad-hoc assumptions and artificial censoring. Sample selection models, on the other hand, allow zero flows and the size of potential trade to be explained jointly and are proved to be the most reasonable choice.

In particular \citet{helpman2008estimating} provide a theoretical framework jointly determining both the set of trading partners and their trade volumes, using the H2S selection model \citep{Heckman1979}. They develop a trade model in which firms face fixed and variable costs of exporting and productivity varies depending on both firms and destinations. Furthermore, trade channels depend on the profitability. Therefore for any pair of countries there may be no firm productive enough to profitably export. As a result, the model is consistent with zero trade flows in both directions between some countries, as well as positive, though asymmetric, trade flows in both directions between others.

Following this literature we carry out all the econometric analyses in this paper using the H2S, which involves first a probit model to estimate the probability of a positive trade flow between any pair of countries and a second step that estimates the log-linear specification 
of the gravity equation on the positive-flow observations, with a selection correction.\footnote{Many alternative models have been introduced in the literature that employ the two-step techniques to address the issue of zero flows. Among them the most common ones are the zero inflated Poisson pseudo maximum likelihood (ZIPPML) and the zero inflated negative binomial maximum likelihood (ZINBML) as well as exponential conditional expectations (ECE), although there has been a long debate in the literature about the appropriateness of different models \citep{silva2015testing,martin2015estimating}.}

\section{Results}\label{sec:results}

In our baseline model we consider the logarithm\footnote{We use the natural logarithm, although the regressions can be run with other bases such as $\log_{10}$. The only difference is that dummy variables coefficients are reduced approximately to half of the value. But all significance levels are unchanged.} 
of directed bilateral trade flows ($\ln trade$) as the dependent variable and the logarithm of the number of direct corporate control links ($\ln CC$) and the logarithm of the shortest path length ($\ln spl$), as a proxy of indirect FDI, as the key explanatory variables. 
In addition, the model includes  
the logarithm of geographical distance (the great-circle definition, $\ln dist$), and the traditional gravity dummy variables including contiguity ($contig$), colony relations ($colony$), whether they have ever been unified ($smcrtry$), 
and common language (if spoken by at least 9\% of the population, $comlang$). Table \ref{tab:summary} shows the summary statistics for the variables used in the regressions. 
Note that we rescale the communicability ($cmb$) values because of their large magnitude and that 90\% of the shortest paths available are indirect\footnote{This justifies our use of the shortest path length as a measure of the indirect effects. If most of the shortest paths are rather direct, the variable $\ln spl$ will be redundant since we already have the direct corporate control intensity, $\ln CC$.}.  

Table \ref{tab:corr} reports the correlation coefficients among the variables used in the regressions. 
Note that $\ln CC$ is correlated with both $\ln spl$ (negatively) and $\ln cmb$ (positively). 
Therefore, we either control $\ln CC$ and $\ln spl$ at the same time to account for the direct and indirect effects of FDI respectively or, as a robustness check, control $\ln cmb$ only to have a comprehensive measure of the FDI effects. 

The results of the baseline models are presented in Table \ref{tab:res_ups_dist_spl}. We have three specifications to account for FDI, $\ln CC$ only,
$\ln spl$ only, or both $\ln CC$ and $\ln spl$. 

To estimate the first step of the H2S selection model, i.e., the probability that a dyad trade relationship exists (extensive margin), we use $\ln dist$\footnote{As far as we know, a unanimous way to model the extensive margin still does not exist.} as the explanatory variable. 
As expected, we find that the probability is negatively correlated with the geographical distance between the two countries. 
In the second step, all the explanatory variables' coefficients have the expected signs and significance, i.e., $\ln gdp$ and $\ln pop$ both have positive and significant coefficients, $\ln dist$ has a negative and significant coefficient, and all the dummies have positive and significant coefficients.\footnote{Our results are robust to additional controls such as common religion, common colonial ties, and landlocking effects.}

Most importantly, we find that the number of corporate control links, $\ln CC$, has a positive and significant effect in explaining trade and the shortest path length, $\ln spl$, has a negative effect in explaining trade. That is, two countries trade more both if they have more direct corporate control links and if they are closer to each other by an indirect path on the CCN.\footnote{The shortest path can also be a direct one. But as stated above, 90\% of time the shortest paths are indirect.} 

Note that our choice of $\ln spl$ rather than $spl$ provides an interesting interpretation of the result. 
If all the shortest paths of the CCN are the direct ones and when $\alpha=1$, $\ln spl$ would
be simply $\ln \frac{1}{CC}=-\ln CC$, according to Equations \ref{eqn:indirPath} and
\ref{eqn:spl}. As a result, the coefficients estimated for either $\ln CC$ or
$\ln spl$ alone would be of the same magnitude but with the opposite 
sign. However, as stated above, about 90\% of the shortest paths are
indirect. Therefore, by comparing the magnitudes of the coefficients
estimated between $\ln CC$ and $\ln spl$, we learn that the indirect
effects are slightly larger than the direct ones. 

We consider also the case when corporate control and trade are in different directions. The right panel of Table \ref{tab:res_ups_dist_spl} shows the regression result by replacing $\ln CC$ with $\ln CC\_inv$ (i.e., the trade importer country controls affiliates in the trade exporter country) and
$\ln spl$ with $\ln spl\_inv$. With this model specification the absolute values of the estimated coefficients of corporate control ($\ln CC\_inv$) and of the shortest path length ($\ln spl\_inv$) both increase, meaning that the number of inverse corporate control links and the inverse shortest path length have larger effects on trade.

As an alternative way to capture the indirect effects, we measure the difference between the shortest path 
and the direct path as $\textit{diff} = \frac{1}{spl} - CC$, where $\frac{1}{spl}$ can be interpreted as the ``potential'' magnititude of FDI stock
of the shortest path, $CC$ is the observed magnititude of FDI stock of 
the direct path, and $\textit{diff}$ can be interpreted as the ``net
gain'' of the shortest path when compared with the direct one. 
Note that, if the shortest path is the direct one, $\textit{diff}=0$, and
if the shortest path is an indirect one, $\textit{diff}>0$. 
Therefore, by our definition above, $\textit{diff}\geq 0$. 
The correlation coefficient between $\ln \textit{diff}$ and $\ln CC$ is 0.466,  which
is much lower in absolute value than that between $\ln spl$ and $\ln CC$, -0.716 (see Table
\ref{tab:corr}).\footnote{Similarly, the correlation coefficient between $\ln \textit{diff}\_inv$ and $\ln CC$ has been lowered to 0.355. The precise
calculation is $\ln \textit{diff} = \ln (\textit{diff}+1)$ and we
add 1 before taking the natural logarithm because $\textit{diff}$ may equal 0 if the shortest path
is the direct one.}
In Table \ref{tab:diff}, we replace $\ln spl$ with $\ln \textit{diff}$ (or
replace $\ln spl\_inv$ with $\ln \textit{diff}\_inv$)
and rerun the baseline models and find that the coefficient of $\ln \textit{diff}$ (or $\ln \textit{diff}\_inv$) is positive and significant. 
This implies that, with other things held constant, more trade is expected
if the ``net gain'' of the shortest path is larger. 

A drawback of our previous analysis is that we measure the strength
of FDI between countries by counting the number of corporate control links,
which may differ from each other in terms of firm and investment sizes.
We mitigate this problem by using an alternative data source of stock FDI from the UNCTAD (United Nations Conference on Trade and Development).\footnote{The stock FDI data set is downloaded from the UNCTAD's Bilateral
FDI Statistics for the year 2010.}
Unlike our measure, the alternative data source has the exact magnitude\footnote{Sometimes the numbers reported by the FDI origin and host countries may be different. In these cases, we take the average.} 
of stock FDI in millions of US dollars. 
We recompute the shortest path length variable based on the UNCTAD data set and report the regression results\footnote{We use the Poisson
pseudo maximum likelihood (PPML) method because the self-selection bias
is absent if using the H2S method.} in Table \ref{tab:unctad}. As before,
a negative and significant coefficient of $\ln spl$ is estimated. Note that
the UNCTAD data set renders much fewer observations than our previous
data set does, which explains why we prefer the number of corporate
controls computed on the firm-level
ORBIS database as a proxy of stock FDI.

Furthermore, in the first column of both panels of Table \ref{tab:res_region_spl} we introduce the interactions of $\ln CC$ and $\ln spl$ with $\ln dist$ into our baseline specification. 
We find that $\ln spl$ flips its sign after the interaction terms are introduced. 
Both $\ln CC$ and $\ln spl$ have opposite signs with respect to their interaction terms. 
Note that the indirect effects of FDI on trade increase with distance while the direct effects of FDI on trade decrease with distance. 
Hence, it is possible to identify the critical value of $dist$ for which $\ln CC$ (or $\ln spl$) starts to affect trade negatively. To do so, we rewrite the model equation as
\begin{equation}\label{eq:easymodel}
	\begin{split}
	\ln trade = \beta_0 + \beta_1 \ln CC + \beta_2 \ln spl + \beta_3 \ln dist \\ + \beta_4 \ln CC\ln dist + \beta_5 \ln spl\ln dist + \ldots
\end{split}
\end{equation}
where, for simplicity, ``\ldots'' indicates other possible factors. The value of $dist$, $\overline{dist}$, we look for is the one that solves
	$\frac{\partial\ln trade}{\partial\ln CC} = \beta_1 + \beta_4 \ln\overline{dist} = 0$ 
	or $\frac{\partial\ln trade}{\partial\ln spl} = \beta_2 + \beta_5 \ln\overline{dist} = 0$. 

For the first column of the left panel of Table \ref{tab:res_ups_dist_spl}, $\overline{dist}$ is about 90335 kilometers for $\ln CC$ and is 2474 kilometers for $\ln spl$. 
In the regression sample, no country pair has distance larger than 90335 kilometers and 11.83\% of the country pairs have distances smaller than 2474 kilometers. 
Therefore, for the majority of country pairs, trade benefits from corporate control relationships, both directly and indirectly.

We also explore other factors of explaining trade. In the second column of both panels of Table \ref{tab:res_region_spl} we analyze the influence of regional trade agreements (RTAs). We are interested in knowing if belonging to a common RTA fosters trade and whether it affects the relation between trade and corporate control. As expected, the presence of RTA ($rta$) has a positive and significant coefficient. 
However, both $\ln CC$ and $\ln spl$ have opposite signs with respect to their interaction terms with $rta$, 
meaning that the presence of RTAs reduces the positive (both direct and indirect) effects of corporate control on trade. This result makes intuitive sense because trade depends less on corporate control links once an RTA, which is composed of explicit arrangements to encourage trade, enters into force. The same result also holds for the inverse direction (the right panel).

Next we shift our attention to Asia. Some
Asian countries may need special treatment as they are active participants of GVCs \citep{baldwin2008spoke,zhu2014rise}. Therefore, we consider the 10 ASEAN countries\footnote{These are Brunei, Cambodia, Indonesia, Laos, Malaysia, Myanmar, Philippines, Singapore, Thailand, and Vietnam.} plus China and add a dummy, $ASEAN+China$, identifying them as trade exporters as well as the interactions of $\ln CC$ and $\ln spl$ with $ASEAN+China$.  
The third column of both panels of Table \ref{tab:res_region_spl} shows the regression result. $ASEAN+China$ has a postive and significant coefficient in both directions while the only significant interaction term is with $\ln CC$ and when both trade and corporate control are in the same direction, meaning that the 11 Asian countries considered carry out more exports than the average and export more to the countries where they have controlled affiliates. 

We also explicitly consider the heterogeneity of the indirect effects across sectors. 
To test this, we exploit the sectoral information of our data set\footnote{We thank the anonymous reviewer
for the suggestion to study the sectoral heterogeneity.} and 
decompose it into six 2-digit NAICS (North American Industry
Classification System) sectors.\footnote{They are: 11, Agriculture, Forestry, Fishing and Hunting; 21, Mining, Quarrying, and Oil and Gas Extraction;
	22, Utilities; and 31-33, Manufacturing. The detailed
information of how we convert HS96 to NAICS is available upon request.}
Table \ref{tab:sectorall} shows the result, where we control for both
the intercept effects by introducing sector dummies and the slope effects
by interacting sector dummies with the number of corporate control
links (i.e., $\ln CC$ in the second column) and the shortest path length variable 
(i.e., $\ln spl$ in the third column), where the benchmark NAICS sector
is 21, Mining, Quarrying, and Oil and Gas Extraction.\footnote{We also run the regressions
	separately for every restricted sample for each sector. The result
is very similar and is available upon request.}
We find that, the manufacturing sectors (31-33) not only have larger intercept effects but also, as expected, have more pronounced indirect effects
of FDI on trade, when compared with the primary sectors (11, 21, and 22).
For example, the sector-specific coefficient of $\ln spl$ is insignificant or positive for 
the primary sectors (e.g., $-0.004+0.091=0.087$ for 11) and turns negative for the manufacturing sectors (e.g., $-0.004-0.268=-0.272$ for 31).   

\section{Robustness Checks}\label{sec:checks}

In this section, we perform a number of robustness checks. 
First of all, there is an endogeneity issue that a mutual
causal linkage may exist between FDI and trade.\footnote{We thank the anonymous
reviewer for raising this point.} Also, Both trade and FDI can be studied in a gravity
model with common determinants such as distance and colonial ties \citep{kleinert2010gravity}. 
To address this issue, we specify the following simultaneous
equations model\footnote{We remove $colony$ from the first equation
	and remove $smctry$ from the second equation to satisfy the 
identification condition.} and run a 3SLS (three-stage least squares) regression \citep{mitze2010trade,wooldridge2015introductory}: 
\begin{equation}
\begin{cases}
	\begin{split}
	 \ln trade = \beta_0 + \beta_1\ln CC + \beta_2\ln spl + \beta_3\ln gdp\_o + \beta_4\ln gdp\_d \\ + 
	 \beta_5\ln pop\_o + \beta_6\ln pop\_d + \beta_7\ln dist \\ + \beta_8 contig + \beta_9 smctry + \beta_{10} comlang + \epsilon  
\end{split}\\ 
\begin{split}
\ln CC = \gamma_0 + \gamma_1\ln trade + \gamma_2\ln gdp\_o + \gamma_3\ln gdp\_d \\ + 
\gamma_4\ln pop\_o + \gamma_5\ln pop\_d + \gamma_6\ln dist \\ + \gamma_7 contig + \gamma_8 colony + \gamma_9 comlang + \upsilon
\end{split}
\end{cases}
\end{equation}
where the endogenous variables are $\ln trade$ and $\ln CC$ and all the
other variables are assumed to be exogenous. 

The reduced form result\footnote{The original form result is available
upon request.} is reported in Table \ref{tab:redsem}. Again, the 
coefficient of $\ln spl$ is significant and negative, which is in line with
our previous results.

Another potential issue with our measure is that aggregating the number of corporate control links across countries may overestimate the strength of indirect control. 
For example, if some firms in country $a$ control some affiliates in country $b$ and some firms in country $b$ have some affiliates in country $c$, we assume that there is an indirect control from country $a$ to country $c$. 
However, country $a$'s affiliates in country $b$ are not necessarily the same ones controlling the affiliates in country $c$.  
Therefore, we need to discount the importance of indirect links with respect to direct ones. 
To do so, we recompute the shortest path length with $\alpha=0.5$.
As a result, 84.2\% of the shortest paths are indirect, as opposed to 90\% if $\alpha=1$. 
However, the main regression results are still robust with $\alpha=0.5$ (see Tables \ref{tab:res_ups_dist_newAlpha} and \ref{tab:res_rta_newAlpha} in the appendix).\footnote{The result of the interaction term between $ASEAN+China$ and $\ln CC$ is not robust with $\alpha=0.5$ (see Table \ref{tab:res_rta_newAlpha} in the appendix).}

As another robustness check, we run the regressions by replacing $\ln CC$ and $\ln spl$ with the overall measure of FDI ``communication'' between countries, the communicability, $\ln cmb$. The main result stays the same and is reported in the appendix (Table \ref{tab:res_ups_dist_com}). 

Last but not least, we also estimate the baseline models using the Poisson pseudo maximum likelihood (PPML) and the zero inflated Poisson pseudo maximum likelihood (ZIPPML). Results are reported in the appendix (Table \ref{tab:checks_spl}) and the signs of the estimated coefficients are indeed robust with respect to different estimation methods.

\section{Conclusions}\label{sec:conclusions}
In this paper we investigate the effects of FDI on trade from a network perspective. We use a unique data set of international corporate control as a measure of stock FDI. We first construct the networks of trade (WTW) and corporate control (CCN) and find a significant correlation between them. 
Most importantly, firms' incentive to reduce tax burden, to minimize 
coordination costs, and to break barriers to market entry, allows the indirect effects of FDI on trade between countries. 

Within the H2S gravity model, we either complement the direct corporate control intensity with the shortest path length or substitute it with the communicability to have a comprehensive accounting of the effects of FDI on trade.
We find that in general corporate control (as a measure of stock FDI) has a positive effect on trade both directly and indirectly. This result is robust with respect to different specifications and estimation strategies, 
therefore providing strong empirical evidence of the indirect effects of FDI on trade. We also identify a number of interplaying factors, including regional trade agreements (RTAs) and the region of Asia.
Moreover, we find that the indirect effects are more pronounced for 
manufacturing sectors than for primary sectors such as oil extraction and 
agriculture.

To extend our work, we may consider the heterogeneity of the corporate control links in the future if more firm-level data become available. Currently we weight the edges of the CCN simply by counting the number of corporate control links between countries. This may be problematic if the links are of very different importance in terms of, for example, economic size.  
Another potential improvement with more firm-level information is that the real indirect corporate control paths can be traced out if we focus on the same firms in the intermediate countries. 
Finally, our work provides strong evidence that indirect FDI is trade-promoting. 
However, indirect FDI comes at a price as it may be the result of tax
evasion. Therefore, the pros and cons of indirect FDI can be
properly evaluated only if more detailed firm-level information is 
obtained.

%


\bibliographystyle{plainnat}
\bibliography{myrefs_ZZ}

\newpage \clearpage

\begin{table}[!t]
  \centering
  \caption{Summary statistics for the variables used in the regressions.}
   	\begin{adjustbox}{width=\textwidth,totalheight=\textheight,keepaspectratio}
    \begin{tabular}{llrrrrr}
    \hline
    Variable & Description  & \# of Observations & Mean  & Std. Dev. & Min   & Max \\
    \hline
    $trade$ & Value of trade in thousands of current USD & 37,442 &  372,915.9  &  4,169,913 & 0     & 328,000,000 \\
    $CC$& Number of corporate control links & 37,442 &  10.17 & 160.97 & 0     & 20,711 \\
    $gdp$  & GDP per capita in thousands of current USD & 36,477 & 11.79    & 16.92 & 0     & 88.41 \\
    $pop$ & Population  & 36,477 & 36,200,000 & 135,000,000 & 20,470 & 1,340,000,000 \\
    $contig$ & 1 if the two countries are contiguous & 36,672 & .0153 & .123  & 0     & 1 \\
    $comlang$ & 1 if the two countries have a common ethnical language & 36,672 & .153  & .360  & 0     & 1 \\
    $colony$ & 1 if the two countries have a colonial relation & 36,672 & .0107 & .103  & 0     & 1 \\
    $smctry$ & 1 if the two countries were/are the same country & 36,672 & .008  & .089  & 0     & 1 \\
    $dist$ & Distance, great circle formula, most important cities/agglomerations & 36,672 & 8,075.14 & 4,591.68 & 1.047.89 & 19,904.45 \\
    $rta$ & 1 if the two countries have a regional trade agreement in force & 35,532  & .069  & .253  & 0     & 1 \\
    $ASEAN+China$ & 1 if the exporter country is an ASEAN country or China &  37,442   &  .056701 &   .2312735         & 0    &      1 \\
    $spl$ & Shortest path length (weighted, directed) based on the CCN & 23,936 & .344  & .452  & .00005  & 3.834 \\
    $cmb$ & Communicability (unweighted, undirected) based on the CCN & 37,442 & .265  & .302   & .0004  & 2.042 \\
    \hline
    \end{tabular}%
    	\end{adjustbox}
  \label{tab:summary}%
\end{table}%

\begin{table}[!t]
	\centering
	\caption{Matrix of the pairwise correlations among the variables used in the
		regressions.}
    \begin{adjustbox}{width=\textwidth,totalheight=\textheight,keepaspectratio}
	\begin{tabular}{lrrrrrrrrrrrrrr}
		\hline
		& $\ln trade$ & $\ln CC$ & $\ln CC\_inv$ & $\ln gdp$ & $\ln pop$ & $comlang$ & $contig$ & $smctry$ & $colony$ & $\ln dist$ & $rta$   & $\ln cmd$ & $\ln spl$ & $\ln spl\_inv$ \\
		\hline
		$\ln CC$ & 0.547 &       &       &       &       &       &       &       &       &       &       &       &       &  \\
		$\ln CC\_inv$ & 0.516 & 0.570 &       &       &       &       &       &       &       &       &       &       &       &  \\
		$\ln gdp$ & 0.295 & 0.445 & 0.176 &       &       &       &       &       &       &       &       &       &       &  \\
		$\ln pop$ & 0.413 & 0.174 & 0.138 & -0.228 &       &       &       &       &       &       &       &       &       &  \\
		$comlang$ & 0.017 & 0.088 & 0.085 & -0.027 & -0.065 &       &       &       &       &       &       &       &       &  \\
		$contig$ & 0.189 & 0.174 & 0.172 & 0.007 & 0.046 & 0.110 &       &       &       &       &       &       &       &  \\
		$smctry$ & 0.088 & 0.065 & 0.066 & -0.007 & -0.019 & 0.104 & 0.311 &       &       &       &       &       &       &  \\
		$colony$ & 0.137 & 0.195 & 0.192 & 0.045 & 0.042 & 0.174 & 0.130 & 0.059 &       &       &       &       &       &  \\
		$\ln dist$ & -0.266 & -0.227 & -0.224 & -0.089 & 0.089 & -0.087 & -0.350 & -0.275 & -0.071 &       &       &       &       &  \\
		$rta$   & 0.283 & 0.303 & 0.302 & 0.169 & -0.061 & 0.022 & 0.189 & 0.141 & 0.063 & -0.555 &       &       &       &  \\
		$\ln cmd$ & 0.652 & 0.594 & 0.597 & 0.436 & 0.218 & -0.044 & 0.044 & -0.020 & 0.086 & -0.020 & 0.216 &       &       &  \\
		$\ln spl$ & -0.506 & -0.716 & -0.434 & -0.649 & -0.131 & -0.004 & -0.051 & 0.000 & -0.061 & 0.019 & -0.210 & -0.722 &       &  \\
		$\ln spl\_inv$ & -0.437 & -0.433 & -0.717 & -0.138 & -0.079 & -0.002 & -0.049 & -0.004 & -0.059 & 0.018 & -0.208 & -0.728 & 0.446 &  \\
		$ASEAN+China$ & 0.156 & 0.000 & 0.041 & -0.122 & 0.283 & -0.037 & -0.004 & -0.016 & -0.023 & 0.143 & -0.078 & 0.070 & -0.100 & -0.038 \\
		\hline
	\end{tabular}%
\end{adjustbox}
	\label{tab:corr}%
\end{table}%

\begin{table}[!t]
	\centering
	\caption{Regressions for the baseline models: (1) $CC$ only; (2) $spl$ only; (3) $CC$ and $spl$. The left panel reports the cases where $trade$ and $CC$ (or $spl$) are in the same direction. The right panel reports the cases where $trade$ and $CC$ (or $spl$) are in the opposite directions. Note that the network measure of the indirect effects is $spl$.} 
\label{tab:res_ups_dist_spl}
	\begin{adjustbox}{width=12cm,keepaspectratio}
	\begin{tabular}{lccc|ccc}
	\hline
	& Baseline (1) &        Baseline (2) &       Baseline (3) &        Baseline (inv) (1) &        Baseline (inv) (2) &        Baseline (inv) (3)   \\
	\hline
	$\ln trade$ (\textbf{Dep. Var.}) &       &       &       &       &       &         \\
	&       &       &       &       &       &        \\
	$\ln CC$ & 0.230  ***   &              & 0.103  ***   &              &              &         \\
	& (0.024) &              &        (0.018) &             &              &                \\
	$\ln CC\_inv$ &              &             &              & 0.276  ***   &              & 0.127  *** \\
	&              &              &              & (0.023)        &              & (0.026)   \\
	$\ln spl$ &              & -0.251  ***   & -0.203  ***   &              &              &         \\
	&             & (0.013)        & (0.016) &              &             &                \\
	$\ln spl\_inv$ &             &             &              &              & -0.306  ***   & -0.246  *** \\
	&              &             &              &              & (0.020)        & (0.022)   \\
	$\ln gdp\_o$ & 1.158  ***   & 0.933  ***   & 0.924  ***   & 1.190  ***   & 1.171  ***   & 1.160  *** \\
	& (0.018)&        (0.017)        & (0.017)        & (0.016)       & (0.018)       & (0.017)   \\
	$\ln gdp\_d$ & 0.832  ***   & 0.817  ***   & 0.809  ***   & 0.773  ***   & 0.604  ***   & 0.594  *** \\
	& (0.016) &        (0.012)        & (0.012)        & (0.017)       & (0.025)        & (0.024)   \\
	$\ln pop\_o$ & 1.097  ***   & 1.059  ***   & 1.047 ***   & 1.100  ***   & 1.110  ***   & 1.099  *** \\
	& (0.013)        & (0.010)        & (0.010)        & (0.012)        & (0.013)        & (0.013)   \\
	$\ln pop\_d$ & 0.862  ***   & 0.870  ***   & 0.861  ***   & 0.845  ***   & 0.831 ***   & 0.818  *** \\
	& (0.012)        & (0.009)        & (0.009)        & (0.012)        & (0.014)      & (0.014)   \\
	$contig$ & 0.566  ***   & 0.429  ***   & 0.399 ***   & 0.554  ***   & 0.417  *     & 0.378  * \\
	& (0.214)       & (0.140)        & (0.139)        & (0.210)       & (0.232)        & (0.223)   \\
	$colony$ & 0.536  ***   & 0.679  ***   & 0.568  ***   & 0.501  ***   & 0.809  ***   & 0.673  *** \\
	& (0.184)       & (0.125)       & (0.126)      & (0.180)       & (0.189)       & (0.184)   \\
	$smctry$ & 1.083  ***   & 0.801  ***   & 0.816  ***   & 1.080  **    & 0.767  **    & 0.787  ** \\
	& (0.295)       & (0.207)       & (0.205)       & (0.289)        & (0.349)        & (0.335)  \\
	$comlang$ & 0.837  ***   & 0.810  ***   & 0.782  ***   & 0.825  ***   & 0.835  ***   & 0.800  *** \\
	& (0.066)        & (0.047)       & (0.047)        & (0.065)        & (0.070)        & (0.067)   \\
	$\ln dist$ & -2.115  ***   & -1.608 ***   & -1.543  ***   & -2.084  ***   & -2.152  ***   & -2.070  *** \\
	& (0.276)        & (0.208)       & (0.206)        & (0.270)        & (0.352)        & (0.339)   \\
	Cons. & -11.694  ***   & -13.626 ***   & -13.667 ***   & -11.659 ***   & -11.547 ***   & -11.615  *** \\
	& (1.602)       & (1.133)       & (1.121)        & (1.568)        & (1.918)        & (1.843)   \\
	\hline
	$trade\_dummy$ (\textbf{Dep. Var.}) &       &       &       &       &       &        \\
	&       &       &       &       &       &        \\
	$\ln dist$ & -0.400  ***   & -0.384  ***   & -0.384  ***   & -0.400  ***   & -0.386  ***   & -0.386  *** \\
	& (0.010) &        (0.010)        & (0.010)        & (0.010)       & (0.010)        & (0.010)   \\
	Cons. & 3.792  ***   & 3.479 ***   & 3.479  ***   & 3.792  ***   & 3.477  ***   & 3.477  *** \\
	& (0.087)       & (0.091)       & (0.091)        & (0.087)       & (0.091)        & (0.091)   \\
	lambda & 4.809  ***   & 1.720  *     & 1.578  *     & 4.708  ***   & 4.625  ***   & 4.443  *** \\
	& (1.313)        & (0.961)       & (0.952)        & (1.285)        & (1.598)        & (1.536)   \\
	&       &       &       &       &       &         \\
	\hline
	N     & 34433 &        29516       & 29516        & 34433       & 29020 &        29020   \\
	\hline
	  	\multicolumn{7}{l}{\footnotesize Standard errors in parentheses; \sym{*} \(p<0.1\), \sym{**} \(p<0.05\), \sym{***} \(p<0.01\).}\\
	\end{tabular}%
\end{adjustbox}	
\end{table}%

\begin{table}[!t]
	\centering
\caption{Poisson pseudo maximum likelihood (PPML) estimates with bilateral stock FDI data from the UNCTAD. Note that we exclude all the negative FDI values before calculating the centrality.}
	\label{tab:unctad}%
	\begin{adjustbox}{width=8cm,keepaspectratio}
	\begin{tabular}{lccc}
		\hline
	& $FDI\_stock$ (1) &        $FDI\_stock$ (2) &        $FDI\_stock$ (3)   \\
	\hline
	$\ln trade$ (\textbf{Dep. Var.}) &       &       &        \\
	&       &       &        \\
	$\ln FDI\_stock$ & 0.156  ***   &              & 0.140  *** \\
	& (0.000)        &              & (0.000)   \\
	$\ln spl$ &              & -0.174  ***   & -0.060  *** \\
	&             & (0.000)        & (0.000)   \\
	$\ln gdp\_o$ & 0.307  ***   & 0.519  ***   & 0.267  *** \\
	& (0.000)        & (0.000)        & (0.000)   \\
	$\ln gdp\_d$ & 0.612  ***   & 0.696  ***   & 0.589  *** \\
	& (0.000) &        (0.000) &        (0.000)   \\
	$\ln pop\_o$ & 0.647  ***   & 0.740  ***   & 0.626  *** \\
	& (0.000) &        (0.000) &        (0.000)   \\
	$\ln pop\_d$ & 0.644  ***   & 0.733  ***   & 0.632  *** \\
	& (0.000) &        (0.000) &        (0.000)   \\
	$contig$ & 0.438  ***   & 0.372  ***   & 0.425  *** \\
	& (0.000) &        (0.000) &        (0.000)   \\
	$colony$ & -0.260  ***   & -0.142  ***   & -0.254  *** \\
	& (0.000) &        (0.000)        & (0.000)   \\
	$smctry$ & 0.532  ***   & 0.633  ***   & 0.496  *** \\
	& (0.000) &        (0.000) &        (0.000)   \\
	$comlang$ & 0.213  ***   & 0.310  ***   & 0.218  *** \\
	& (0.000) &        (0.000) &        (0.000)   \\
	$\ln dist$ & -0.463  ***   & -0.568  ***   & -0.463  *** \\
	& (0.000) &        (0.000) &        (0.000)   \\
	Cons.  & -7.544  ***   & -11.400  ***   & -7.205  *** \\
	& (0.000) &       (0.000) &        (0.000)   \\
	&       &       &         \\
\hline
	N     & 4810         & 4810 &       4810    \\
		\hline
\multicolumn{4}{l}{\footnotesize Standard errors in parentheses; \sym{*} \(p<0.1\), \sym{**} \(p<0.05\), \sym{***} \(p<0.01\).}\\
	\end{tabular}%
\end{adjustbox}	
\end{table}%

\begin{table}[!t]
	\centering
\caption{Regressions with $dist$, $rta$, and $ASEAN+China$, with the interactions of $CC$ and $spl$ with $dist$, $rta$, and $ASEAN+China$. The left panel reports the cases where $trade$ and $CC$ (or $spl$) are in the same direction. The right panel reports the cases where $trade$ and $CC$ (or $spl$) are in the opposite directions. Note that the network measure of the indirect effects is $spl$.}
	\label{tab:res_region_spl}
	\begin{adjustbox}{width=12cm,keepaspectratio}
	\begin{tabular}{lccc|ccc}
		\hline
		& $dist$  &        $rta$         & $ASEAN+China$ &        $dist$ (inv) &        $rta$ (inv) &        $ASEAN+China$ (inv)   \\
		\hline
		$\ln trade$ (\textbf{Dep. Var.}) &       &       &       &       &       &         \\
		&       &       &       &       &       &         \\
		$\ln dist$ & -1.899  ***   & -1.201  ***   & -1.400  ***   & -2.460  ***   & -1.682  ***   & -1.886  *** \\
		& (0.209) &       (0.207) &        (0.195) &        (0.367) &        (0.279)        & (0.257)   \\
		$\ln CC$ & 0.702  ***   & 0.167  ***   & 0.128  ***   &              &             &         \\
		& (0.170)        & (0.021)       & (0.018)        &              &             &         \\
		$\ln CC\_inv$ &            &             &              & 0.340        & 0.198  ***   & 0.116  *** \\
		&              &              &              & (0.275)        & (0.022)        & (0.020)   \\
		$(\ln dist)\times(\ln CC)$ & -0.061 ***   &              &              &              &             &         \\
		& (0.020)       &              &       &       &       &         \\
		$(\ln dist)\times(\ln CC\_inv)$ &             &              &              & -0.018        &              &         \\
		&             &             &              & (0.031)        &              &        \\
		$\ln spl$ & 1.580  ***   & -0.215  ***   & -0.144  ***   &              &              &         \\
		& (0.160)        & (0.016)       & (0.016)        &             &              &        \\
		$\ln spl\_inv$ &              &              &              &    1.054***         &   -.255  ***            & -0.243  *** \\
		&       &       &       &  (0.255)     &    (0.018)   &       (0.017)   \\
		$(\ln dist)\times(\ln spl)$ & -0.202  ***   &              &              & &             &         \\
		& (0.018)       &       &                    & &              &         \\
		$(\ln dist)\times(\ln spl\_inv)$ &              &              &              & -0.147***        &              &        \\
		&              &              &             & (0.029)        &              &         \\
		$rta$   &              & 1.676***        &              &              & 1.880  ***   &         \\
		&              & (0.112)        &              &              & (0.136)        &         \\
		$rta\times(\ln CC)$ &             & -0.022        &              &              &              &         \\
		&              & (0.043)        &              &              &              &         \\
		$rta\times(\ln CC\_inv)$ &              &              &              &              & -0.045        &         \\
		&             &              &              &              & (0.051)        &         \\
		$rta\times(\ln spl)$ &              & 0.339***        &              &              &              &         \\
		&              & (0.042)        &              &              &              &         \\
		$rta\times(\ln spl\_inv)$ &              &              &              &              & 0.096  **    &         \\
		&              &              &              &             & (0.049)        &        \\
		$ASEAN+China$ &              &              & 1.599  ***   &              &              & 1.265  *** \\
		&              &              & (0.137)        &              &              & (0.135)   \\
		$(ASEAN+China)\times(\ln spl)$ &              &              & 0.051        &              &              &         \\
		&              &              & (0.048)        &              &              &         \\
		$(ASEAN+China)\times(\ln CC)$ &              &              & 0.149  **    &             &              &         \\
		&              &             & (0.063)        &              &              &         \\
		$(ASEAN+China)\times(\ln spl\_inv)$ &              &              &              &              &              & -0.018   \\
		&              &              &              &             &              & (0.054)   \\
		$(ASEAN+China)\times(\ln CC\_inv)$ &              &              &              &              &              & -0.016   \\
		&              &             &              &              &              & (0.066)   \\
		$\ln gdp\_o$ & 0.929  ***   & 0.927  ***   & 0.983  ***   & 1.166  ***   & 1.145  ***   & 1.169  *** \\
		& (0.017)       & (0.017)        & (0.017)        & (0.018)        & (0.014)        & (0.013)   \\
		$\ln gdp\_d$ & 0.817  ***   & 0.796  ***   & 0.825  ***   & 0.595  ***   & 0.593  ***   & 0.600  *** \\
		& (0.012)       & (0.012)       & (0.012)       & (0.025)       & (0.019)       & (0.018)   \\
		$\ln pop\_o$ & 1.041 ***   & 1.046  ***   & 1.002  ***   & 1.095  ***   & 1.093  ***   & 1.067  *** \\
		& (0.010) &        (0.010) &        (0.010) &        (0.013) &        (0.010) &        (0.010)   \\
		$\ln pop\_d$ & 0.858  ***   & 0.856  ***   & 0.873  ***   & 0.811  ***   & 0.815  ***   & 0.823  *** \\
		& (0.009) &        (0.009) &        (0.009) &        (0.014)       & (0.011)       & (0.011)   \\
		$contig$ & 0.398  ***   & 0.424 ***   & 0.343  **    & 0.408  *     & 0.430  **    & 0.349  ** \\
		& (0.136) &        (0.135)        & (0.131)       & (0.231)       & (0.175)        & (0.169)   \\
		$colony$ & 0.544  ***   & 0.479  ***   & 0.604  ***   & 0.644  ***   & 0.572  ***   & 0.714  *** \\
		& (0.125) &        (0.125)        & (0.123)        & (0.189)       & (0.142)        & (0.140)   \\
		$smctry$ & 0.643  ***   & 0.554  ***   & 0.801  **    & 0.662  *     & 0.514  **    & 0.794  *** \\
		& (0.197) &        (0.194)        & (0.191)       & (0.345)        & (0.260)        & (0.254)  \\
		$comlang$ & 0.704  ***   & 0.715  ***   & 0.801  ***   & 0.730  ***   & 0.735  ***   & 0.825  *** \\
		& (0.047) &        (0.047)       & (0.046)        & (0.069)        & (0.052)        & (0.051)  \\
		Cons. & -10.084  ***   & -15.926  ***   & -13.867  ***   & -8.155  ***   & -14.236  ***   & -12.089  *** \\
		& (1.144) &        (1.122)       & (1.055)        & (2.006)        & (1.511)        & (1.398)   \\
		\hline
		$trade\_dummy$ (\textbf{Dep. Var.}) &       &       &       &       &       &       \\
		&       &       &       &       &       &         \\
		$\ln dist$ & -0.384  ***   & -0.378  ***   & -0.384  ***   & -0.386  ***   & -0.380  ***   & -0.386  *** \\
		& (0.010) &        (0.010)       & (0.010)        & (0.010)        & (0.010)       & (0.010)   \\
		Cons. & 3.479  ***   & 3.417  ***   & 3.479  ***   & 3.477  ***   & 3.415  ***   & 3.477  *** \\
		& (0.091) &       (0.091)        & (0.091)        & (0.091)      & (0.092)        & (0.091)   \\
		lambda & 1.099 &        0.607        & 0.712        & 4.568  ***   & 3.435 ***   & 3.339  *** \\
		& (0.981) &        (0.959)        & (0.900)        &(1.679)        & (1.261)       & (1.167)   \\
		&       &       &       &       &       &        \\
		\hline
		N     & 29516 &        29193 &        29516 &        29020 &        28711 &       29020  \\
		\hline
			\multicolumn{7}{l}{\footnotesize Standard errors in parentheses; \sym{*} \(p<0.1\), \sym{**} \(p<0.05\), \sym{***} \(p<0.01\).}\\
	\end{tabular}%
\end{adjustbox}	
\end{table}%

\begin{table}[!t]
	\centering
	\caption{Heckman two step model with sector fixed effects and the interactions of sector dummies (denoted by 2-digit numbers) with $CC$ and $spl$. The benchmark NAICS sector is 21, Mining, Quarrying, and Oil and Gas Extraction.}
	\label{tab:sectorall}%
	\begin{adjustbox}{width=5cm,keepaspectratio}
	\begin{tabular}{lcc}
		\hline
		& $CC$   &        $spl$     \\
		\hline
		$\ln trade$ (\textbf{Dep. Var.}) &       &         \\
		& & \\
		$[sectors]$ &       &        \\
		11 (Agr.)    & 1.388  ***   & 1.615  *** \\
	& (0.043) &        (0.061)   \\
	22 (Uti.)    & -0.438  ***   & -0.161  ** \\
		& (0.047)       & (0.066)   \\
			31 (Man.1)    & 2.746  ***   & 2.899  *** \\
			& (0.041) &        (0.058)   \\
		32 (Man.2)    & 2.732  ***   & 2.800  *** \\
		& (0.041) &        (0.058)   \\
		33 (Man.3)    & 2.847  ***   & 2.958  *** \\
		& (0.041)       & (0.057)   \\
		$\ln CC$ & -0.002     & \\
		& (0.051) &        \\
		$\ln spl$ &       &        -0.004   \\
		&       &       (0.017)   \\
		$[sectors\times(\ln CC)]$ &       &        \\
		11 (Agr.)    & -0.141      &         \\
		& (0.115)        &         \\
		22 (Uti.)    & 0.157  *   &         \\
		& (0.083)        &         \\
		31 (Man.1)    & 0.453  ***   &         \\
		& (0.067)        &         \\
		32 (Man.2)    & 0.665  ***   &         \\
		& (0.060)        &         \\
		33 (Man.3)    & 0.771  ***   &         \\
		& (0.058)        &         \\
		$[sectors\times(\ln spl)]$ &              &         \\
		11 (Agr.)    &              & 0.091  *** \\
		&             & (0.021)   \\
		22 (Uti.)    &             & -0.047  ** \\
		&              & (0.022)   \\
		31 (Man.1)    &              & -0.268  *** \\
		&              & (0.020)   \\
		32 (Man.2)    &             & -0.430  *** \\
		&       &        (0.020)   \\
		33 (Man.3)    &       &        -0.507  *** \\
		&       &        (0.020)   \\
		&       &                \\
		$\ln gdp\_o$ & 0.956  ***   & 0.647  *** \\
		& (0.007) &        (0.011)   \\
		$\ln gdp\_d$ & 0.715  ***   & 0.753  *** \\
		& (0.007) &        (0.007)   \\
		$\ln pop\_o$ & 0.960  ***   & 0.913  *** \\
		& (0.006) &        (0.006)   \\
		$\ln pop\_d$ & 0.726  ***   & 0.764  *** \\
		& (0.005) &        (0.005)   \\
		$contig$ & 0.878  ***   & 0.719  *** \\
		& (0.079) &        (0.068)   \\
		$colony$ & 0.605  ***   & 0.793  *** \\
		& (0.068) &        (0.061)   \\
		$smctry$ & 0.677  ***   & 0.329  *** \\
		& (0.113) &        (0.102)   \\
		$comlang$ & 0.765  ***   & 0.692  *** \\
		& (0.028) &        (0.026)   \\
		$\ln dist$ & -2.257  ***   & -1.494  *** \\
		& (0.177) &        (0.197)  \\
		Cons. & -11.067  ***   & -13.830  *** \\
		& (0.851) &        (0.794)   \\
		\hline
		$trade\_dummy$ (\textbf{Dep. Var.}) &       &        \\
		&       &      \\
		$\ln dist$ & -0.379  ***   & -0.341***   \\
		& (0.004) &        (0.004)   \\
		Cons. & 3.048 ***   & 2.518***   \\
		& (0.032) &        (0.035)   \\
		lambda & 4.819  ***   & 1.236***   \\
		& (0.718) &        (0.834)   \\
		&       &              \\
		\hline
		N     & 213031 &        188935   \\
		\hline
			\multicolumn{3}{l}{\footnotesize Standard errors in parentheses; \sym{*} \(p<0.1\), \sym{**} \(p<0.05\), \sym{***} \(p<0.01\).}\\
	\end{tabular}%
\end{adjustbox}	
\end{table}%

\appendix

\setcounter{table}{0}
\renewcommand\thetable{A\arabic{table}}

\setcounter{figure}{0}
\renewcommand\thefigure{A\arabic{figure}}


\clearpage

\begin{table}[!t]
	\centering
	\caption{Regressions for the baseline model using $\ln \textit{diff}$ to replace $\ln spl$. The second column reports the case where $trade$ and $CC$ (or $\textit{diff}$) are in the same direction. The third column reports the case where $trade$ and $CC$ (or $\textit{diff}$) are in the opposite directions.}
	\label{tab:diff}%
	\begin{adjustbox}{width=5cm,keepaspectratio}
	\begin{tabular}{lcc}
		\hline
		& $\textit{diff}$ &     $\textit{diff}$ (inv)         \\
		\hline
		$\ln trade$ (\textbf{Dep. Var.}) &       &                    \\
		&       &                      \\
		$\ln CC$ & 0.182  ***   &                         \\
		& (0.016) &                                  \\
		$\ln CC\_inv$ &              & 0.223 ***                          \\
		&              &       (0.026)                       \\
		$\ln \textit{diff}$ &   0.175  ***           &                \\
		&    (0.014)         &                       \\
		$\ln \textit{diff}\_inv$ &             &  0.209***                         \\
		&              &     (0.023)                         \\
		$\ln gdp\_o$ & 0.947  ***   & 1.166  ***      \\
		& (0.016)&        (0.020)                  \\
		$\ln gdp\_d$ & 0.813  ***   & 0.625  ***      \\
		& (0.012) &        (0.026)                 \\
		$\ln pop\_o$ & 1.053  ***   & 1.102  ***       \\
		& (0.010)        & (0.015)                  \\
		$\ln pop\_d$ & 0.863  ***   & 0.825  ***      \\
		& (0.009)        & (0.016)                 \\
		$contig$ & 0.483  ***   & 0.477  ***      \\
		& (0.145)       & (0.254)                  \\
		$colony$ & 0.637  ***   & 0.751  ***        \\
		& (0.129)       & (0.211)                \\
		$smctry$ & 0.834  ***   & 0.816  ***        \\
		& (0.214)       & (0.382)                \\
		$comlang$ & 0.801  ***   & 0.823  ***      \\
		& (0.047)        & (0.077)                  \\
		$\ln dist$ & -1.649  ***   & -2.198 ***     \\
		& (0.214)        & (0.386)                 \\
		Cons. & -13.309  ***   & -11.198 ***      \\
		& (1.167)       & (2.103)                 \\
		\hline
		$trade\_dummy$ (\textbf{Dep. Var.}) &       &                \\
		&       &                    \\
		$\ln dist$ & -0.384  ***   & -0.386  ***      \\
		& (0.010) &        (0.010)                  \\
		Cons. & 3.479  ***   & 3.477 ***      \\
		& (0.091)       & (0.091)                 \\
		lambda & 2.102  **   & 5.069  ***          \\
		& (0.986)        & (1.749)                 \\
		&       &                     \\
		\hline
		N     & 29516 &        29020              \\
		\hline
			\multicolumn{3}{l}{\footnotesize Standard errors in parentheses; \sym{*} \(p<0.1\), \sym{**} \(p<0.05\), \sym{***} \(p<0.01\).}\\
	\end{tabular}%
\end{adjustbox}
\end{table}%

\begin{table}[!t]
	\centering
	\caption{Simultaneous equations model (SEM) with three-stage least-squares regression. The reduced model is reported.}
\label{tab:redsem}%
	\begin{adjustbox}{width=5cm,keepaspectratio}
	\begin{tabular}{lcc}
		\hline
		&  $\ln trade$ (\textbf{Dep. Var.}) &               $\ln CC$ (\textbf{Dep. Var.})   \\
		    \hline
		    $\ln spl$ & -0.218  ***   &         \\
		    & (0.013)       &         \\
		    $\ln gdp\_o$ & 0.962  ***   & 0.499  *** \\
		    & (0.017)        & (0.007)   \\
		    $\ln\_gdp\_d$ & 0.832  ***   & 0.270  *** \\
		    & (0.012) &        (0.006)   \\
		    $\ln\_pop\_o$ & 1.067  ***   & 0.236  *** \\
		    & (0.010) &        (0.005)   \\
		    $\ln\_pop\_d$ & 0.877  ***   & 0.190  *** \\
		    & (0.009) &        (0.004)   \\
		    $contig$ & 0.515 *** &    0.861  *** \\
		    & (0.126) &        (0.065)   \\
		    $colony$ & 0.662  ***&    0.516  *** \\
		    & (0.124) &        (0.066)   \\
		    $smctry$ & 0.906  *** &   0.029   \\
		    & (0.185) &        (0.097)   \\
		    $comlang$ & 0.821  ***   & 0.410  *** \\
		    & (0.046) &        (0.024)   \\
		    $\ln dist$ & -1.237  ***   & -0.310 *** \\
		    & (0.021) &        (0.011)   \\
		    Cons. & -15.862  ***   & -5.228  *** \\
		    & (0.303) &        (0.146)   \\
		    &       &         \\
		   \hline
		    N     & 15870 & 15870        \\
		\hline
\multicolumn{3}{p{0.8\textwidth}}{\footnotesize Endogenous  variables: $\ln trade$ and $\ln CC$; Exogenous  variables: $\ln spl$, $\ln gdp\_o$,  $\ln gdp\_d$,  $\ln pop\_o$, $\ln pop\_d$, $contig$, $colony$, $smctry$,  $comlang$, and $\ln dist$. Standard errors in parentheses; \sym{*} \(p<0.1\), \sym{**} \(p<0.05\), \sym{***} \(p<0.01\).}\\
	\end{tabular}%
\end{adjustbox}	
\end{table}%

\begin{table}[!t]
	\centering
\caption{Regressions for the baseline models: (1) $CC$ only; (2) $spl$ only; (3) $CC$ and $spl$. The left panel reports the cases where $trade$ and $CC$ (or $spl$) are in the same direction. The right panel reports the cases where $trade$ and $CC$ (or $spl$) are in the opposite directions. Note that the network measure of the indirect effects is $spl$, which is computed with $\alpha=0.5$ in Equation \ref{eqn:indirPath}.}
	\label{tab:res_ups_dist_newAlpha}
	\begin{adjustbox}{width=12cm,keepaspectratio}
	\begin{tabular}{lccc|ccc}
		\hline
		      & Baseline (1) &        Baseline (2) &        Baseline (3) &        Baseline (inv) (1) &        Baseline (inv) (2) &        Baseline (inv) (3)   \\
		      \hline
			  $\ln trade$ (\textbf{Dep. Var.}) &       &       &       &       &       &        \\
		      &       &       &       &       &       &         \\
		      $\ln CC$ & 0.230  ***   &              & 0.079  ***   &              &              &        \\
		      & (0.024)        &              & (0.018)        &              &              &        \\
		      $\ln CC\_inv$ &             &             &              & 0.276  ***   &              & 0.074  *** \\
		      &              &              &              & (0.023)        &              & (0.025)   \\
		      $\ln spl$ &              & -0.563  ***   & -0.480  ***   &              &              &         \\
		      &              & (0.028)        & (0.034)        &              &             &         \\
		      $\ln spl\_inv$ &             &              &             &             & -0.734  ***   & -0.655  *** \\
		      &              &              &              &              & (0.039)        & (0.045)   \\
		      $\ln gdp\_o$ & 1.158  ***   & 0.900  ***   & 0.897  ***   & 1.190  ***   & 1.138  ***   & 1.135  *** \\
		      & (0.018)       & (0.018)       & (0.018)        & (0.016)        & (0.017)       & (0.017)   \\
		      $\ln gdp\_d$ & 0.832 ***   & 0.798 ***   & 0.794 ***   & 0.773 ***   & 0.544 ***   & 0.543 *** \\
		      & (0.158) &        (0.012) &        (0.012) &        (0.017) &        (0.024) &        (0.024)   \\
		      $\ln pop\_o$ & 1.097 ***   & 1.044 ***   & 1.037 ***   & 1.100 ***   & 1.091 ***   & 1.086 *** \\
		      & (0.013) &        (0.010) &        (0.010) &        (0.012) &       (0.013) &       (0.012)   \\
		      $\ln pop\_d$ & 0.862 ***   & 0.859 ***   & 0.853 ***   & 0.845 ***   & 0.808 ***   & 0.802 *** \\
		      & (0.012) &        (0.009) &        (0.010) &        (0.012) &        (0.013) &        (0.013)   \\
		      $contig$ & 0.566 ***   & 0.397 ***   & 0.378 ***   & 0.554 ***   & 0.371  *     & 0.353  * \\
		      & (0.214) &        (0.138) &        (0.138) &        (0.210) &        (0.214) &        (0.210)   \\
		      $colony$ & 0.536 ***   & 0.648 ***   & 0.568 ***   & 0.501 ***   & 0.777 ***   & 0.701 *** \\
		      & (0.184) &        (0.124) &        (0.126) &        (0.180) &        (0.174) &        (0.173)   \\
		      $smctry$ & 1.084 ***   & 0.807 ***   & 0.818 ***   & 1.080 ***   & 0.771  **    & 0.782  ** \\
		      & (0.295) &        (0.204) &        (0.202) &        (0.289) &        (0.323) &        (0.316)   \\
		      $comlang$ & 0.837 ***   & 0.769 ***   & 0.753 ***   & 0.825 ***   & 0.779 ***   & 0.764 *** \\
		      & (0.066) &        (0.047) &        (0.047) &        (0.065) &        (0.065) &        (0.064)   \\
		      $\ln dist$ & -2.115 ***   & -1.568 ***   & -1.526 ***   & -2.084 ***   & -2.082 ***   & -2.038 *** \\
		      & (0.276) &       (0.205) &        (0.204) &        (0.270) &        (0.326) &        (0.320)   \\
		      Cons. & -11.694 ***   & -13.211 ***   & -13.304 ***   & -11.659 ***   & -10.963 ***   & -11.064 *** \\
		      & (1.602) &       (1.116) &        (1.110) &        (1.568) &        (1.774) &        (1.740)   \\
		     \hline
			 $trade\_dummy$ (\textbf{Dep. Var.}) &       &       &       &       &       &       \\
		      &       &       &       &       &       &        \\
		      $\ln dist$ & -0.400 ***   & -0.384 ***   & -0.384 ***   & -0.400 ***   & -0.386 ***   & -0.386 *** \\
		      & (0.010) &        (0.010)        & (0.010) &        (0.010) &        (0.010) &        (0.010)   \\
		      Cons. & 3.792 ***   & 3.479 ***   & 4.479 ***   & 3.792 ***   & 3.478 ***   & 3.477 *** \\
		      & (0.087) &        (0.091) &       (0.090) &        (0.087) &        (0.091) &        (0.091)   \\
		      lambda & 4.808 ***   & 1.523        & 1.444        & 4.708 ***   & 4.278 ***   & 4.195 *** \\
		      & (1.313) &        (0.948) &        (0.944) &        (1.285) &        (1.478) &       (1.450)   \\
		      &       &       &       &       &       &        \\
		    \hline
		      N     & 34433 &        29516        & 29516       & 34433        & 29020        & 29020  \\
		\hline
\multicolumn{7}{l}{\footnotesize Standard errors in parentheses; \sym{*} \(p<0.1\), \sym{**} \(p<0.05\), \sym{***} \(p<0.01\).}\\
	\end{tabular}%
\end{adjustbox}
\end{table}%

\begin{table}[!t]
	\centering
\caption{Regressions with $dist$, $rta$, and $ASEAN+China$, with the interactions of $CC$ and $spl$ with $dist$, $rta$, and $ASEAN+China$. The left panel reports the cases where $trade$ and $CC$ (or $spl$) are in the same direction. The right panel reports the cases where $trade$ and $CC$ (or $spl$) are in the opposite directions. Note that the network measure of the indirect effects is $spl$, which is computed with $\alpha=0.5$ in Equation \ref{eqn:indirPath}.}
	\label{tab:res_rta_newAlpha}
	\begin{adjustbox}{width=12cm,keepaspectratio}
	\begin{tabular}{lccc|ccc}
		\hline
		& $dist$  &        $rta$   &        $ASEAN+China$ &        $dist$ (inv) &        $rta$ (inv) &        $ASEAN+China$ (inv)   \\
		\hline
		$\ln trade$ (\textbf{Dep. Var.}) &       &       &       &       &       &         \\
		&       &       &       &       &       &         \\
		$\ln dist$ & -1.726  ***   & -1.168  ***   & -1.381  ***   & -2.353  ***   & -1.651  ***   & -1.855  *** \\
		& (0.206) &        (0.206) &        (0.194) &        (0.351) &        (0.269) &       (0.251)   \\
		$\ln CC$ & 1.027  ***   & 0.140  ***   & 0.111  ***   &       &       &         \\
		& (0.183)        & (0.021)        & (0.019)        &       &       &         \\
		$\ln CC\_inv$ &              &              &              & 0.376        & 0.145  ***   & 0.063  *** \\
		&                    &      &       &        (0.284) &        (0.022)        & (0.020)   \\
		$(\ln dist)\times(\ln CC)$ & -0.101  ***   &              &              &              &              &         \\
		& (0.021)        &       &       &       &       &        \\
		$(\ln dist)\times(\ln CC\_inv)$ &              &              &              & -0.028        &             &         \\
		&             &              &              & (0.032)        &              &         \\
		$\ln spl$ & 3.709  ***   & -0.508  ***   & -0.342  ***   &              &              &         \\
		& (0.340)        & (0.035)       & (0.035)        &              &              &         \\
		$\ln spl\_inv$ &              &              &              & 2.070  ***   & -0.674  ***   & -0.651  *** \\
		&             &             &              & (0.521)        & (0.037)        & (0.037)   \\
		$(\ln dist)\times(\ln spl)$ & -0.476  ***   &              &              &              &              &         \\
		& (0.038)        &       &       &       &       &        \\
		$(\ln dist)\times(\ln spl\_inv)$ &              &              &              & -0.310  ***   &              &         \\
		&             &              &              & (0.058)        &              &         \\
		$rta$   &              & 1.557  ***   &              &              & 1.711  ***   &         \\
		&              & (0.096)        &              &              & (0.115)        &         \\
		$rta\times(\ln CC)$ &              & 0.047        &              &              &              &         \\
		&              & (0.047)        &              &              &              &         \\
		$rta\times(\ln CC\_inv)$ &              &             &              &              & 0.008        &        \\
		&              &              &              &             & (0.055)        &         \\
		$rta\times(\ln spl)$ &              & 0.814  ***   &              &              &              &         \\
		&              & (0.092)        &              &              &             &         \\
		$rta\times(\ln spl\_inv)$ &              &              &              &              & 0.120        &         \\
		&             &              &              &              & (0.105)       &        \\
		$ASEAN+China$ &             &             & 1.437  ***   &              &             & 1.253 *** \\
		&              &              & (0.124)        &              &              & (0.114)   \\
		$(ASEAN+China)\times(\ln spl)$ &              &              & -0.024        &              &              &         \\
		&              &              & (0.109)        &              &              &         \\
		$(ASEAN+China)\times(\ln CC)$ &              &              & 0.105        &              &              &         \\
		&              &              & (0.067)        &              &             &         \\
		$(ASEAN+China)\times(\ln spl\_inv)$ &              &              &              &              &              & -0.025   \\
		&              &              &              &              &              & (0.067)   \\
		$(ASEAN+China)\times(\ln CC\_inv)$ &              &              &              &              &              & -0.066   \\
		&              &             &              &              &              & (0.114)   \\
		$\ln gdp\_o$ & 0.902 ***   & 0.902 ***   & 0.962 ***   & 1.140 ***   & 1.122 ***   & 1.143 *** \\
		& (0.018) &        (0.018) &        (0.018) &        (0.017) &        (0.014)        & (0.013)   \\
		$\ln gdp\_d$ & 0.804 ***   & 0.784 ***   & 0.813 ***   & 0.542 ***   & 0.543 ***   & 0.548 *** \\
		& (0.012) &        (0.012) &        (0.012) &        (0.024) &        (0.019) &        (0.019)   \\
		$\ln pop\_o$ & 1.030 ***   & 1.036 ***   & (0.995 ***   & 1.082 ***   & 1.081 ***   & 1.054 *** \\
		& (0.010) &        (0.010) &        (0.010) &        (0.012) &        (0.010) &        (0.010)   \\
		$\ln pop\_d$ & 0.851 ***   & 0.849 ***   & 0.866 ***   & 0.795 ***   & 0.800 ***   & 0.807 *** \\
		& (0.009) &        (0.009) &        (0.009) &        (0.014) &        (0.011) &        (0.011)   \\
		$contig$ & 0.411 ***   & 0.417 ***   & 0.331 **    & 0.419 *     & 0.423 **    & 0.324 ** \\
		& (0.133) &        (0.134) &        (0.131) &        (0.219) &        (0.169) &        (0.165)   \\
		$colony$ & 0.555 ***   & 0.486 ***   & 0.606 ***   & 0.678 ***   & 0.610 ***   & 0.744 *** \\
		& (0.124) &        (0.124) &        (0.123) &        (0.179) &        (0.139) &        (0.138)  \\
		$smctry$ & 0.642 ***   & 0.531 ***   & 0.802 ***   & 0.664 **    & 0.498 **    & 0.789 *** \\
		& (0.193) &        (0.193) &        (0.190) &        (0.328) &        (0.251) &        (0.247)   \\
		$comlang$ & 0.670 ***   & 0.688 ***   & 0.779 ***   & 0.692 ***   & 0.702 ***   & 0.789 *** \\
		& (0.047) &        (0.047) &        (0.046) &        (0.066) &        (0.051) &        (0.051)   \\
		Cons. & -10.848 ***   & -15.675 ***   & -13.607 ***   & -8.276 ***   & -13.711 ***   & -11.530 *** \\
		& (1.126) &        (1.116) &        (1.050) &        (1.916)&        (1.460) &        (1.362)   \\
		\hline
		$trade\_dummy$ (\textbf{Dep. Var.}) &       &       &       &       &       &        \\
		&       &       &       &       &       &        \\
		$\ln dist$ & -0.384 ***   & -0.378 ***   & -0.384 ***   & -0.386 ***   & -0.380 ***   & -0.386 *** \\
		& (0.010) &        (0.010) &        (0.010) &        (0.010) &        (0.010) &        (0.010)   \\
		Cons. & 3.479 ***   & 3.418 ***   & 3.479 ***   & 4.478 ***   & 3.415 ***   & 3.478 *** \\
		& (0.091) &        (0.091) &        (0.091) &       (0.091) &        (0.092) &        (0.091)   \\
		lambda & 0.637        & 0.417        & 0.586        & 4.341 ***   & 3.189 ***   & 3.092 *** \\
		& (0.964) &       (0.953) &        (0.897) &       (1.600) &        (1.218) &        (1.138)   \\
		&       &       &       &       &       &        \\
		\hline
		N     & 29516 &        29193 &        29516 &       29020 &        28711 &        29020  \\
		\hline
\multicolumn{7}{l}{\footnotesize Standard errors in parentheses; \sym{*} \(p<0.1\), \sym{**} \(p<0.05\), \sym{***} \(p<0.01\).}\\
	\end{tabular}%
\end{adjustbox}
\end{table}%

\begin{table}[!t]
	\centering
 	\caption{Regressions for the baseline models and for the specifications with the interactions of $cmb$ with $dist$, $rta$, and $ASEAN+China$. Note that the network measure of the indirect effects is $cmb$. }
 	\label{tab:res_ups_dist_com}
 	\begin{adjustbox}{width=10cm,keepaspectratio}
	\begin{tabular}{lcccc}
	\hline
	& Baseline &        $dist$  &        $rta$   &        $ASEAN+China$ \\
	\hline
	$\ln trade$ (\textbf{Dep. Var.}) &       &       &       &         \\
	& & & & \\
	$\ln cmd$ & 0.852  ***   & -1.432  ***   & 0.839  ***   & 0.811  *** \\
	& (0.042) &        (0.225)        & (0.029)        & (0.033)   \\
	$\ln dist$ & -2.234  ***   & -1.739 ***   & -1.768  ***   & -2.064  *** \\
	& (0.284) &        (0.246)      & (0.205)        & (0.223)   \\
	$(\ln dist)\times(\ln cmd)$ &             & 0.269  ***   &              &         \\
	&             & (0.026)        &              &         \\
	$rta$   &              &              & 0.110        &         \\
	&             &              & (0.082)        &         \\
	$ASEAN+China$ &              &              &              & 1.771  *** \\
	&              &              &              & (0.117)   \\
	$rta\times(\ln cmd)$ &              &              & -0.690  ***   &         \\
	&              &              & (0.059)        &         \\
	$(ASEAN+China)\times(\ln cmd)$ &             &             &              & 0.372  *** \\
	&              &              &              & (0.080)   \\
	$\ln gdp\_o$ & 0.950  ***   & 0.960  ***   & 0.976  ***   & 0.969  *** \\
	& (0.022) &       (0.018) &        (0.015) &        (0.017)   \\
	$\ln gdp\_d$ & 0.581  ***   & 0.589  ***   & 0.605  ***   & 0.587  *** \\
	& (0.021) &        (0.018)        & (0.015)       & (0.017)   \\
	$\ln pop\_o$ & 0.973  ***   & 0.971  ***   & 0.993  ***   & 0.941  *** \\
	& (0.015) &        (0.013) &        (0.010) &        (0.012)   \\
	$\ln pop\_d$ & 0.726  ***   & 0.724  ***   & 0.745  ***   & 0.732  *** \\
	& (0.015) &        (0.013)      & (0.010)        & (0.012)   \\
	$contig$ & 0.665  ***   & 0.754  ***   & 0.610 ***   & 0.628  *** \\
	& (0.222) &        (0.189) &        (0.154) &        (0.173)   \\
	$colony$ & 0.898  ***   & 0.966  ***   & 0.889  ***   & 0.945  *** \\
	& (0.189) &        (0.162) &        (0.129)        & (0.148)  \\
	$smctry$ & 0.976  ***   & 0.763  ***   & 0.557  ***   & 0.971  *** \\
	& (0.305) &        (0.262) &        (0.210) &        (0.238)   \\
	$comlang$ & 0.799  ***   & 0.736  ***   & 0.749  ***   & 0.821  *** \\
	& (0.068) &        (0.059) &        (0.047)        & (0.053)   \\
	Cons. & -4.511  ***   & -8.249  ***   & -8.376  ***   & -5.077 *** \\
	& (1.694) &        (1.483) &        (1.220) &        (1.324)   \\
	\hline
	$trade\_dummy$ (\textbf{Dep. Var.}) &       &       &       &        \\
	& & & & \\
	$\ln dist$ & -0.400  ***   & -0.400  ***   & -0.395  ***   & -0.400  *** \\
	& (0.010) &        (0.010) &        (0.010) &        (0.010)   \\
	Cons. & 3.792  ***   & 3.792  ***   & 3.732  ***   & 3.792  *** \\
	& (0.087) &        (0.087) &        (0.087) &        (0.087)  \\
	lambda & 4.976  ***   & 4.252 ***   & 3.392  ***   & 3.888  *** \\
	& (1.355) &        (1.158)        & (0.969)        & (1.062)   \\
	&       &       &       &         \\
\hline
	N     & 34433        & 34433 &        33991 &        34433   \\
		\hline
\multicolumn{5}{l}{\footnotesize Standard errors in parentheses; \sym{*} \(p<0.1\), \sym{**} \(p<0.05\), \sym{***} \(p<0.01\).}\\
	\end{tabular}%
\end{adjustbox}
\end{table}%

\begin{table}[!t]
	\centering
\caption{Diagnostic checks using the baseline models: (1) $CC$ only; (2) $spl$ only; (3) $CC$ and $spl$. OLS, Poisson pseudo maximum likelihood (PPML) and zero inflated Poisson pseudo maximum likelihood (ZIPPML) are used. The network measure of the indirect effects is $spl$.}
  \label{tab:checks_spl}
	\begin{adjustbox}{width=12cm,keepaspectratio}
	\begin{tabular}{lccc|ccc|ccc}
		\hline
	   & OLS (1) &        OLS (2)        & OLS (3)       & PPML (1)        & PPML (2)        & PPML (3)       & ZIPPML (1)       & ZIPPML (2)       & ZIPPML (3)  \\
	   \hline
	   $\ln trade$ (\textbf{Dep. Var.}) &       &       &       &       &       &       &       &       &      \\
	   & & & & & & & & & \\
	   $\ln CC$ & 0.237  ***   &             & 0.104  ***   & 0.086  ***   &             & 0.114  ***   & 0.089  ***   &            & 0.119  *** \\
	   & (0.016)        &              & (0.018)        & (0.000)        &              & (0.000) &        (0.000) &              &        (0.000)   \\
	   $\ln spl$ &             & -0.253  ***   & -0.205  ***   &              & -0.079  ***   & 0.024  ***   &              & -0.078  ***   & 0.030  *** \\
	   &             & (0.013)        & (0.016)        &              & (0.000)        & (0.000)        &       &      (0.000)        & (0.000)   \\
	   $\ln gdp\_o$ & 1.160 ***   & 0.932 ***   & 0.922 ***   & 0.616 ***   & 0.614 ***   & 0.564 ***   & 0.601 ***   & 0.610 ***   & 0.557 *** \\
	   & (0.012) &        (0.017) &        (0.017) &        (0.000) &        (0.000) &        (0.000) &        (0.000) &        (0.000)        & (0.000)   \\
	   $\ln gdp\_d$ & 0.836 ***   & 0.818 ***   & 0.809 ***   & 0.742 ***   & 0.756 ***   & 0.732 ***   & 0.731 ***   & 0.751 ***   & 0.724 *** \\
	   &(0.011) &       (0.012) &       (0.012) &       (0.000)        &(0.000) &       (0.000) &       (0.000) &       (0.000)        &(0.000)   \\
	   $\ln pop\_o$ & 1.093 ***   & 1.058 ***   & 1.046 ***   & 0.793 ***   & 0.790 ***   & 0.768 ***   & 0.783 ***   & 0.786 ***   & 0.762 *** \\
	   &(0.009) &       (0.010)       &(0.010)        &(0.000)       &(0.000)        &(0.000)        &(0.000)        &(0.000)        &(0.000)   \\
	   $\ln pop\_d$ & 0.858 ***   & 0.869 ***   & 0.860 ***   & 0.754 ***   & 0.767 ***   & 0.741 ***   & 0.745 ***   & 0.762 ***   & 0.735 *** \\
	   &(0.008) &       (0.009) &       (0.009) &       (0.000) &       (0.000) &       (0.000) &       (0.000) &       (0.000) &       (0.000)   \\
	   $contig$ & 0.789 ***   & 0.503 ***   & 0.466 ***   & 0.428 ***   & 0.434 ***   & 0.441 ***   & 0.435 ***   & 0.439 ***   & 0.447 *** \\
	   &(0.117) &       (0.126)        &(0.126)       &(0.000)        &(0.000)      &(0.000)        &(0.000)        &(0.000)        &(0.000)   \\
	   $colony$ & 0.519 ***   & 0.677 ***   & 0.566 ***   & -0.164 ***   & -0.119 ***   & -0.177 ***   & -0.170 ***   & -0.122 ***   & -0.183 *** \\
	   &(0.121) &       (0.124) &       (0.125) &       (0.000) &      (0.000) &       (0.000) &       (0.000) &       (0.000) &       (0.000)   \\
	   $smctry$ & 1.322 ***   & 0.899 ***   & 0.906 ***   & 0.751 ***   & 0.683 ***   & 0.714 ***   & 0.723 ***   & 0.667 ***   & 0.697 *** \\
	   &(0.160) &       (0.185) &       (0.185) &       (0.000) &       (0.000) &       (0.000) &       (0.000) &       (0.000) &       (0.000)   \\
	   $comlang$ & 0.844 ***   & 0.810 ***   & 0.782 ***   & 0.249 ***   & 0.303 ***   & 0.240 ***   & 0.256 ***   & 0.309 ***   & 0.244 *** \\
	   &(0.044) &       (0.046) &       (0.047) &       (0.000) &     (0.000) &       (0.000) &       (0.000) &       (0.000) &       (0.000)   \\
	   $\ln dist$ & -1.123 ***   & -1.240 ***   & -1.204 ***   & -0.567 ***   & -0.596 ***   & -0.554 ***   & -0.565 ***   & -0.595 ***   & -0.551 *** \\
	   &(0.021) &       (0.021) &       (0.022) &       (0.000) &       (0.000) &       (0.000) &       (0.000) &       (0.000)        &(0.000)   \\
	   Cons.  & -17.209 ***   & -15.558 ***   & -15.440 ***   & -11.850 ***   & -11.900 ***   & -11.060 ***   & -11.450 ***   & -11.710 ***   & -10.800 *** \\
	   &(0.272) &     (0.303) &       (0.304) &       (0.000) &       (0.000) &       (0.000) &       (0.000) &       (0.000)        &(0.000)   \\
	  \hline
	  $trade\_dummy$ (\textbf{Dep. Var.}) &       &       &       &       &       &       &       &       &        \\
	  & & & & & & & & & \\
	   Cons.  &       &       &       &       &       &       &        -7.375 ***   & -8.654 ***   & -8.656 *** \\
	   &       &       &       &       &       &       &       (0.177)        &(0.267) &       (0.267)   \\
	   $\ln dist$ &       &       &       &       &       &       &        0.752 ***   & 0.827 ***   & 0.828 *** \\
	   &       &       &       &       &       &       &      (0.020) &       (0.030) &       (0.030)   \\
	   &       &       &       &       &       &       &       &       &        \\
	  \hline
	   N     & 20776       & 15858        & 15858       & 30779        & 20126        & 20126       & 30779        & 20126        & 20126   \\
		\hline
\multicolumn{10}{l}{\footnotesize Standard errors in parentheses; \sym{*} \(p<0.1\), \sym{**} \(p<0.05\), \sym{***} \(p<0.01\).}\\
	\end{tabular}%
\end{adjustbox}	
\end{table}%

\end{document}